\documentclass[journel,10pt,twocolumn]{IEEEtran}

\usepackage{latexsym}
\usepackage{amsfonts}
\usepackage{amssymb}
\usepackage{amsbsy}
\usepackage{amsmath}
\usepackage[dvips]{epsfig}

\usepackage[tight,footnotesize]{subfigure}
\usepackage[dvips]{epsfig}

\begin{document}

\title{Game-theoretic Understanding of Price Dynamics in Mobile Communication Services}

\author{\IEEEauthorblockN{Seung Min Yu and Seong-Lyun
Kim} \thanks{Seung Min Yu and Seong-Lyun Kim are with the School of
Electrical and Electronic Engineering, Yonsei University, 50
Yonsei-Ro, Seodaemun-Gu, Seoul 120-749, Korea. E-mail: \{smyu,
slkim\} @ramo.yonsei.ac.kr. All correspondence should be addressed
to Prof. Seong-Lyun Kim.}}

\maketitle

\begin{abstract}
In the mobile communication services, users wish to subscribe to
high quality service with a low price level, which leads to
competition between mobile network operators (MNOs). The MNOs
compete with each other by service prices after deciding the extent
of investment to improve quality of service (QoS). Unfortunately,
the theoretic backgrounds of price dynamics are not known to us, and
as a result, effective network planning and regulative actions are
hard to make in the competitive market. To explain this competition
more detail, we formulate and solve an optimization problem applying
the two-stage Cournot and Bertrand competition model. Consequently,
we derive a price dynamics that the MNOs increase and decrease their
service prices periodically, which completely explains the subsidy
dynamics in the real world. Moving forward, to avoid this
instability and inefficiency, we suggest a simple regulation rule
which leads to a Pareto-optimal equilibrium point. Moreover, we
suggest regulator's optimal actions corresponding to user welfare
and the regulator's revenue.
\end{abstract}

\begin{keywords}
Network economics, game theory, competition, price dynamics,
regulation, mobile communications.
\end{keywords}

\section{Introduction}

\subsection{Conflict of Interests among Mobile Network Operators, Users, and the Regulator}
\begin{figure}[t]
\centerline{\epsfig{figure=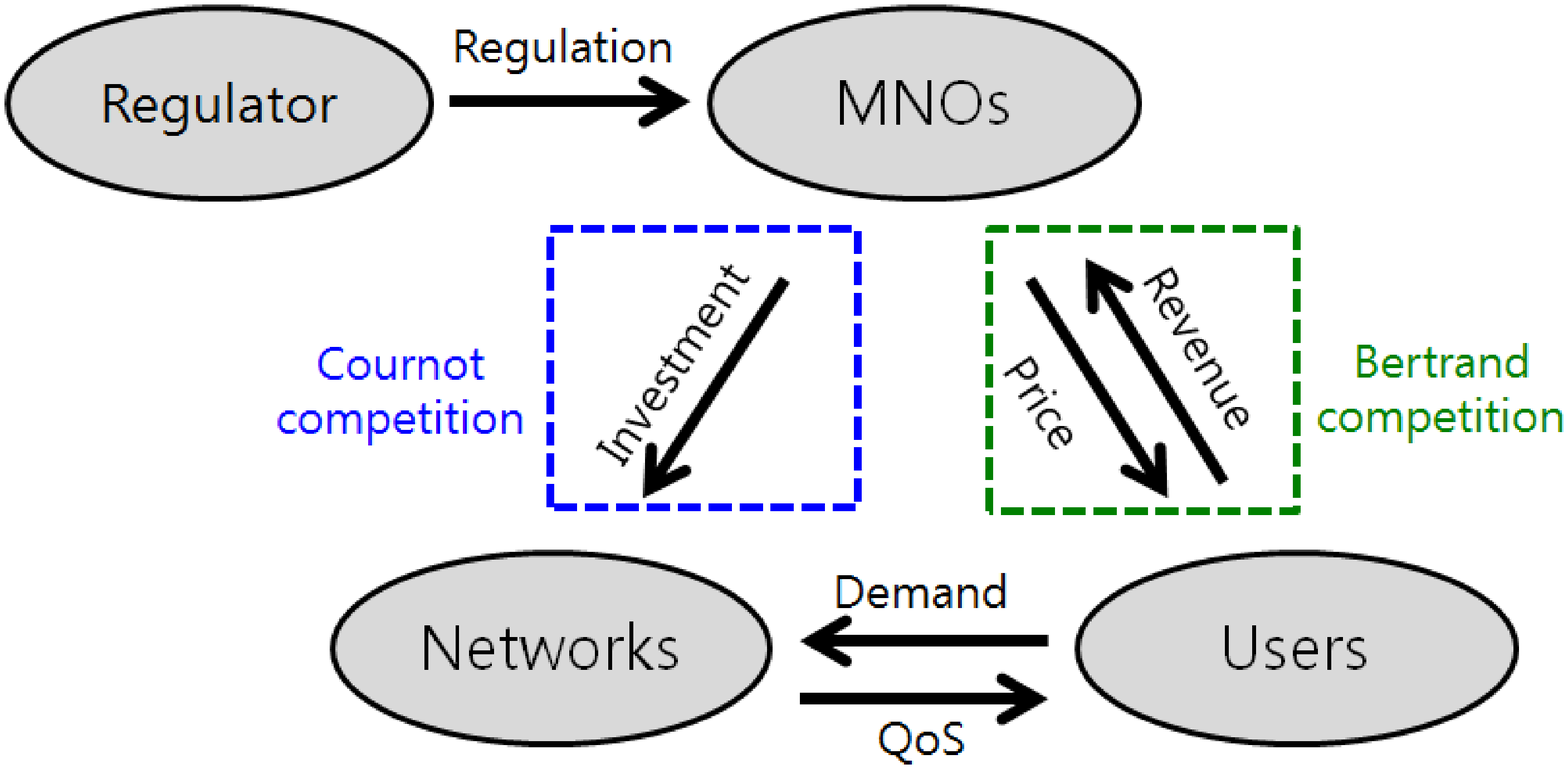,width=3.4in,height=1.7in,clip=;}}
\caption{The interaction among MNOs, users, and the regulator.}
\label{structure}
\end{figure}

In mobile communication services, there is interaction among {\it
mobile network operators (MNOs)}, {\it users}, and {\it the
regulator} (Figure \ref{structure}). Each MNO makes an investment in
its network to improve the quality of service (QoS) and sets a
service price to maximize its profit. The users decide which MNO is
more appropriate to subscribe to the network service considering the
service price and the QoS. Finally, the regulator aims to maximize
the welfare of all users. Therefore, there should be some
equilibrium points for the service price and network investment
(QoS) between MNOs and users. Theoretically, finding such
equilibrium points is not easy. The situation becomes even more
complicated when there are multiple MNOs competing with each other.

Maximizing profit is the primary concern of MNOs, which might be
achieved by having a high price level and low investment on the
network. On the other hand, users wish to maximize their utility by
consuming high QoS with a low service price. The QoS is directly
related to the network investment from MNOs. Therefore, there is a
conflict of interests among these players and the role of the
regulator is very important. From the regulatory perspective, solely
maximizing profit by MNOs should be avoided if it is at the cost of
sacrificing user welfare significantly.

For making efficient regulations, we firstly investigate
characteristics of the competitive mobile communication services. An
important question for MNOs is how much of the network capacity
should be provisioned and how high the service price should be.
Price competition between two operators was previously studied by
Walrand \cite{Walrand}, where the network capacity was assumed to be
given. Here, we analyze how each MNO determines the optimal
investment on the network and the service price as a response to the
strategy of its competitor. For this purpose, we apply Cournot and
Bertrand competition models \cite {Colell}-\cite {Bertrand}.

In the Cournot model, MNOs compete with each other deciding the
extent of investment on their networks. On the other hand, in the
Bertrand model, MNOs engage in price competition to attract more
subscribers for a given network capacity. We combine the Cournot and
Bertrand models so that the network capacity is determined in the
Cournot phase and afterwards the service price is determined in the
Bertrand phase. The Cournot and Bertrand models are interlinked and
we achieve joint optimization of the network capacity and service
price. Our main viewpoint of this joint optimization is in
investigating the dynamics of competition between MNOs and also in
finding an optimal role of the regulator.

\subsection{Price Competition and Subsidization}
The dynamics of price competition among network operators was
studied in some previous works \cite {Yu}-\cite {Kong}.
Particularly, in \cite{Yu}, we showed that there would be a price
dynamics that network operators increase and decrease their service
prices periodically. In the real world, however, network operators'
billing systems are very similar in the same country or state and
the price dynamics does not seem to occur.

For discussing the reality of the price dynamics, let us consider
the monthly charging structure in the mobile communication service.
In most countries, many MNOs give a {\it subsidy} to attract new
subscribers (potential users or their competitors' subscribers)
\cite{Kim}. The subsidy is offered as part of a contract that
includes a stipulated time period. Therefore, we should consider the
subsidy amount for examining the price competition among MNOs.

\begin{figure}[t]
\centerline{\epsfig{figure=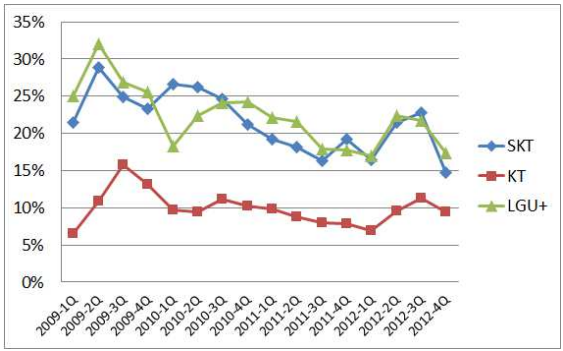,height=1.8in,clip=;}}
\caption{Quarterly marketing expenses as percentage of sales of the
major MNOs (SKT, KT and LGU+) in South Korea.} \label{subsidy}
\end{figure}

To show that there can be a kind of price dynamics by subsidization
in the real world, we plot quarterly marketing expenses as
percentage of sales of the major MNOs (SKT, KT and LGU+) in South
Korea as an example (Figure \ref{subsidy}). Note that the
investigated marketing expenses are mostly used for subsidization.
In the figure, the MNOs increase and decrease their marketing
expenses repeatedly, and this can be interpreted as the service
price dynamics by subsidization. Then, why do the MNOs use
subsidization as an indirect method for increasing or decreasing
their service prices? The MNOs cannot increase the service prices
easily due to regulations. On the other hand, there are few
regulations on subsidization and people are relatively generous
about change of the subsidy amount because it is believed that
subsidization is a means of lowering the cost of new subscriber's
entry to the mobile communication services \cite{Gruber}. Therefore,
the MNOs compete with each other by adjusting their subsidy amounts,
making the price dynamics in the real world. Unfortunately,
theoretic backgrounds of price dynamics are not known to us, and as
a result, effective network planning and regulative actions are hard
to make in the competitive market.

In this paper, we analyze the price dynamics between MNOs using the
two stage competition model, where the MNOs increase and decrease
their service prices periodically without an equilibrium point. This
kind of price dynamics is not desirable to any player due to the
instability. For example, it is unfair that users' payments for the
mobile communication service are different in different start times
of the subscription even though they are served by the same MNO.
Based on our analysis, to avoid such instability, we suggest a
simple regulation rule that guarantees an equilibrium point of price
levels, which is Pareto-optimal.

\subsection{Related Work}
In \cite{Odlyzko}, the author shows that service price and QoS are
inter-related in communications networks, and suggests Paris metro
pricing (PMP) for Internet. PMP is a kind of price discrimination
over different QoS levels; the higher QoS, the higher price. In that
paper, the author finds that the service price and QoS will converge
to an equilibrium point after a number of interactions. PMP is
further extended by Walrand \cite{Walrand}, who formulates an
Internet pricing model under price and QoS constraints. In that
work, the author investigates how much PMP improves the operator's
profit compared to a single optimal service price. The author also
analyzes price competition between two homogeneous network
operators, the network capacities of which are fixed. In \cite{Yu},
we show the dynamics of price competition ({\it price war}) using
the Walrand model \cite{Walrand}, and suggest a regulation for price
level convergence.

The price war in communication service is observed in \cite
{Yu}-\cite {Kong}. Particularly, in \cite{Chiu} and \cite{Tan}, if
one operator lowers its price to increase revenue or to monopolize
the entire market, then the other operators will also lower their
price to match the price leader. The price down competition will
occur repeatedly among all operators, eventually damaging every
operator with a revenue decrease.

Competition among network operators occurs not only by price
differentiation. The capacity of the network is another important
variable. This is because users will select a network operator based
on decision criteria including not only service price but also QoS
level, and the QoS is directly related to the network capacity.
Therefore, each operator jointly optimizes the service price and
network capacity. All of the previous work mentioned above focuses
only on price competition, assuming the network capacity is given as
an external value. In \cite{Shetty}, the authors consider
competition among multiple network operators with single- or
two-service classes. In that work, service prices are fixed and
price competition does not occur. To attract more users, the
operators decide only the network capacity.

Suppliers of a homogeneous good/service compete with each other by
deciding their amount of output. This is called Cournot competition
(quantity competition) \cite{Cournot}. Generally, the market price
decreases as the total amount of output increases. On the other
hand, Bertrand competition refers to price competition where the
suppliers compete with each other by controlling the product price
\cite{Bertrand}. In the Bertrand competition model, consumers buy
all of a particular product from the supplier with the lowest price.

We analyze mobile communications markets using Cournot and Bertrand
competition models \cite{Wpin}-\cite{Jung2}. In
\cite{Wpin}-\cite{Jung1}, we suggest spectrum policies and
subsidization schemes for improving user welfare in mobile
communications. In \cite{Jung2}, we investigate the effect of
allocation of asymmetric-valued spectrum blocks on mobile
communications markets. However, our previous works focus on
spectrum allocation and have not dealt with price dynamics in mobile
communications.

\subsection{Main Contribution of This Paper}
Using the two stage model \cite{Kreps}, we will show that MNOs
sequentially decrease their service prices (i.e., increase
subsidies) as in \cite{Chiu} and \cite{Tan}, but one MNO suddenly
increases its price when the competitor's price is lower than a
certain threshold. Therefore, the price levels increase and decrease
periodically without an equilibrium point. We call this {\it price
war with long jumps}, which is not desirable to any player.

The main contributions and results of this paper are summarized
below.
\begin{itemize}
\item {\it Description of price dynamics}: In the real world, MNOs tend to compete with each other changing their service prices
by subsidization (see Figure \ref{subsidy}). Using a two-stage
Cournot and Bertrand competition model with network congestion, we
mathematically analyze the competition between MNOs. Based on a
game-theoretic approach, we show that there exists price (subsidy)
dynamics in the mobile communication service, which well explains
the subsidy dynamics in Figure \ref{subsidy}.

\item {\it Regulation for price convergence}: To avoid instability and inefficiency, we propose a simple regulation that limits the number of price level
changes. We show that the regulation guarantees an equilibrium point
of price levels that is Pareto-optimal. We also introduce more
realistic regulations that bring the same effect of the price
regulation.

\item {\it Regulation under a two-stage Cournot and Bertand model}: Using the two-stage model, we calculate an
equilibrium point of the network capacity and the service price.
From the result, we suggest a regulator's optimal action (exacting
taxes) corresponding to user welfare or the regulator's revenue.
This is an extension of our previous work \cite{Yu} to the two-stage
model.
\end{itemize}

The rest of this paper is organized as follows. In the next section,
we describe our system model. Section III presents our optimization
problem in two-stage duopoly competition. In Section IV, we derive a
solution to the optimization problem in the Bertrand stage and
explain how two operators' prices vary. We suggest a simple
regulation that drives the price levels to converge on an
equilibrium point. In Section V, we combine the Bertrand model with
the Cournot model. Using a backward induction method, we solve the
optimization problem in the Cournot stage and find an equilibrium
point. From the results, we describe characteristics of the
communications service market and introduce the role of the
regulator. Finally, Section VI concludes the paper.

\section{System Model}
Consider a service area covered by two competitive MNOs. There are
$M$ users for the mobile communication service. Nonnegative values
$k_1$ and $k_2$, respectively, denote the first and the second MNO's
capacity, which determine the quality of service (QoS) of the
networks. MNOs determine the optimal $k_1$ and $k_2$ values in the
Cournot stage. In the Bertrand stage, MNOs compete by controlling
$p_1$ and $p_2$, the first and the second MNO's price for the
service. These service prices include the subsidy amounts (i.e., the
initial service price minus the subsidy amount). Therefore, the MNOs
can control the service prices by adjusting the subsidy amounts even
if there are some regulations that prohibit the initial price level
changes. Without loss of generality, we assume $p_1$ and $p_2$ are
normalized values over the interval $[0,1]$. Each MNO can provide
only one price to all users at a given time. The QoS of a network
depends on the congestion level of the network. We denote the QoS of
each MNO's network by $q_1$ and $q_2$, respectively. Without loss of
generality, the values of $q_1$ and $q_2$ are also normalized over
the interval $[0,1]$. A value closer to 0 denotes a higher
congestion level (lower QoS).

Each user decides whether to subscribe to the communication service
or not by selecting its serving MNO. Some users prefer high QoS (low
congestion) even though they have to pay more. On the other hand,
some other users will accept a low QoS if the service price is low,
as was also noted by Paris metro pricing \cite{Odlyzko}.
Willingness-to-pay and the QoS required by users are positively
correlated. To model user behavior, we define the {\it user type} as
in \cite{Walrand} and \cite{Shetty}. The user type $\alpha$ is a
variable over $[0,1]$ that quantifies the user's willingness-to-pay.
At the same time, it quantifies the QoS level required by the
user.\footnote{The user type $\alpha$ has a dual role as
willingness-to-pay and QoS requirement. This seems to be open to
dispute because those two criteria cannot merged into
one-dimensional parameter space. In this paper, however, we assume
that each user's willingness-to-pay and QoS requirement are highly
correlated and can be modeled as a one-dimensional parameter for the
mathematical tractability.} The value $\alpha$ is close to $1$ when
the user is willing to pay a high price for high QoS. At the other
extreme ($\alpha\xrightarrow{{}}0$), the user prefers low QoS with a
low price. Since it is difficult to figure out the user type value
of each user, we assume that it is a random variable (e.g., uniform
distribution in $[0,1]$).

Consider a user with user type value $\alpha$. For the user to
subscribe to the communication service offered by MNO $i$, both the
price and QoS levels should be satisfied. In other words, $\alpha
\ge p_i$ and $\alpha \le q_i$, where we regard the first and second
inequalities as the {\it price condition} (PC) and {\it QoS
condition} (QC), respectively.\footnote{The characteristics of
users' MNO selection are based on the assumption that each user
wants a specific service (or application) requiring some target QoS
level. Then, each user's utility function becomes a step function
with a step at QoS level. In other words, if a QoS level is higher
than $\alpha$, then the utility from the service is equal to
$\alpha$. Otherwise, the utility is zero. Therefore, a user with
user type $\alpha$ subscribes to MNO $i$'s communication service
only if both $\alpha \ge p_i$ and $\alpha \le q_i$ are satisfied.}
The PC is commonly used in microeconomics \cite{Phillips}, but it is
not sufficient to model communication service, where some users
whose PCs are satisfied may not use the service because the QoS is
lower than expected due to congestion. If multiple MNOs satisfy both
conditions, then the user will select the MNO offering the lowest
price.

We use a linearly decreasing QoS model as in \cite{Walrand} and
\cite{Shetty}, which mirrors the perception of service quality
\cite{Gibbens}:
\begin{eqnarray} \label{eq:QoS}
q_1  = 1 - \frac{{d_1 }} {{k_1 }M},\,\,\,\,\,q_2 = 1 - \frac{{d_2 }}
{{k_2 }M},
\end{eqnarray}
where $d_1$ and $d_2$ denote the number of users accessing the first
and the second MNO, respectively. We define the reference capacity,
which makes the QoS level equal to zero when all users access one of
the MNOs. $k_1$ and $k_2$ are normalized values by the reference
capacity. In other words, if $k_1=1$ ($k_2=1$) and $d_1=M$
($d_2=M$), then the QoS level is $q_1=0$ ($q_2=0$). The user demand
is expressed as follows:
\begin{eqnarray} \label{eq:demand}
d_1  = M\int_{\alpha _1^{\min } }^{\alpha _1^{\max } } f \left(
\alpha \right)d\alpha, \,\,\,\,\,d_2  = M\int_{\alpha _2^{\min }
}^{\alpha _2^{\max } } f \left( \alpha  \right)d\alpha,
\end{eqnarray}
where $\alpha _1^{\min }$, $\alpha _1^{\max }$, $\alpha _2^{\min }$
and $\alpha _2^{\max }$ denote the minimum and the maximum values of
$\alpha$ among the users accessing the first and the second MNO,
respectively. $f \left( \alpha  \right)$ denotes a probability
density function of $\alpha$. Equation (\ref {eq:demand}) is a
demand function derived from integration of the willingness-to-pay
distribution.

\begin{figure}[t]
\centerline{\epsfig{figure=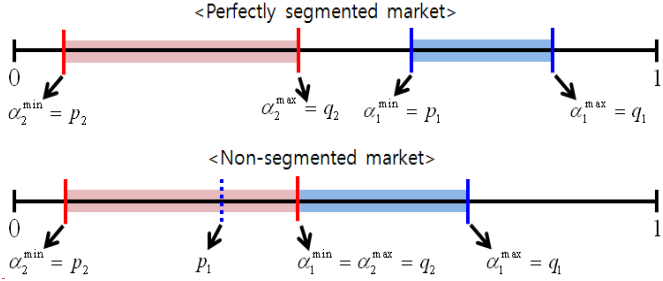,width=3.45in,height=1.5in,clip=;}}
\caption{User type intervals of the users subscribing each MNO's
service.} \label{integral_interval}
\end{figure}

Figure \ref{integral_interval} illustrates a perfectly segmented
market and non-segmented market. Assume $p_1>p_2$. In the perfectly
segmented market ($p_1 \ge q_2$), the values of $\alpha _1^{\min }$
and $\alpha _2^{\min }$ are determined by PC, and $\alpha _1^{\max
}$ and $\alpha _2^{\max }$ are determined by QC. On the other hand,
in the non-segmented duopoly market ($p_1 < q_2$), $\alpha_1^{\min}$
is determined by $\alpha_2^{\max}$. This is because if there are
users whose PC and QC are satisfied by both MNOs\footnote{For the
users in [$p_1,q_2$], both PC and QC are satisfied by both MNOs.},
then the users whose user types are within $[p_1,q_2]$ will access
the second MNO with the lower price, $p_2$.

\section{Mobile Network Operator's Optimization Problem}
\begin{figure}[t]
\centerline{\epsfig{figure=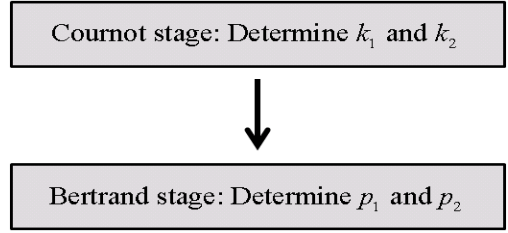,width=2.6in,height=1.2in,clip=;}}
\caption{Two-stage Cournot and Bertrand competition between two
MNOs.} \label{two_stage}
\end{figure}

Figure \ref{two_stage} explains our two-stage model. In the Cournot
stage, MNOs decide their capacity considering the investment cost.
Each MNO cannot change its network capacity in the short-term after
observing a competitor's network investment. Thus, this capacity
competition can be modeled as a simultaneous game. Hereafter, let
$i$ denote the decision maker index and $j$ denote the competitor's.
Then, we formulate the optimization problem of MNO $i$ in the
Cournot stage as follows:
\begin{eqnarray} \label{eq:cournot_opt_problem}
\mathop {{\text{max}}}\limits_{k_i \ge 0} \,\,\,\, f_i^R \left( {k_i
,k_j } \right) - f_i^C \left( {k_i } \right),  \hfill
\end{eqnarray}
where $f_i^R \left(  \cdot  \right)$ and $f_i^C \left(  \cdot
\right)$ denote revenue and cost functions of MNO $i$. Note that the
MNO's revenue depends not only on its own capacity, but also on its
competitor's. The revenue function is determined by the result of
the Bertrand stage.

In the Bertrand stage, MNOs compete with each other by controlling
their prices for the given capacity determined in the Cournot stage.
Here the price includes the subsidy. Thus high price implies low
subsidy, and low price implies high subsidy. Each MNO can change its
price repeatedly after observing a competitor's. Thus, this price
competition can be modeled as an infinite sequential game. We
exclude the case of pricing equal to that of the
competitor.\footnote{In \cite{Shetty}, the authors assume that all
MNOs' service prices are the same due to MNO competition. Thus,
price dynamics does not occur. Each MNO in our model, on the other
hand, has no reason to match its competitor's service price because
it can profit more by lowering its service price a little, which
will eventually lead to a price dynamics.} Then, we formulate the
optimization problem of MNO $i$ in the Bertrand stage, which is
divided into two cases: Using either a higher or lower price than
the competitor's.

\begin{itemize}
\item {\it Using a lower price ($p_i \le p_j$)}:
\begin{eqnarray} \label{eq:lower_problem}
  \mathop {{\text{max}}}\limits_{0 \le p_i \le 1} \,\,\,\,&&p_i d_i  \hfill \nonumber \\
  {\text{s}}{\text{.t}}{\text{.}}\,\,\,\,\,\, &&d_i  = M\int_{p_i }^{1 - \frac{{d_i }}
{k_i M} } f \left( \alpha  \right)d\alpha.  \hfill \nonumber
\end{eqnarray}
\end{itemize}
In this case, the number of users for MNO $i$ is independent of the
competitor's price, which is just like the monopoly led by the low
pricing MNO. Thus, the upper and lower limits of the integral in the
constraint are replaced by $\alpha_i^{\max}=q_i=1-d_i/k_i M$ and
$\alpha_i^{\min}=p_i$ as in Figure \ref{integral_interval}.

\begin{itemize}
\item {\it Using a higher price ($p_i>p_j$)}:
\begin{eqnarray} \label{eq:higher_problem}
  \mathop {{\text{max}}}\limits_{0 \le p_i \le 1} \,\,\,\,&&p_i d_i  \hfill \nonumber \\
  {\text{s}}{\text{.t}}{\text{.}}\,\,\,\,\,\, &&d_i  = M\int_{\max
\left\{ {p_i ,1 - \frac{{d_j }} {k_j M}} \right\}}^{1 - \frac{{d_i
}} {k_i M}} {f \left( \alpha  \right)d\alpha }.  \hfill \nonumber
\end{eqnarray}
\end{itemize}
In this case, the competitor affects the number of users of MNO $i$.
If the competitor guarantees PC and QC of a user, then the user will
access the competitor network. Noting that $\alpha_i^{\max}$ is
determined by the QC, the upper limit of the integral is given by
$\alpha_i^{\max}=q_i=1-d_i/k_i M$. On the other hand,
$\alpha_i^{\min}$ is determined by $\max{\{p_i,q_j\}}$ as explained
in Figure \ref{integral_interval}.

\section{Bertrand Stage: Price (Subsidy) Competition}
A common method for analyzing a multi-stage game is the backward
induction method. This method is used to find the equilibrium that
represents a Nash equilibrium in every stage (or subgame) of the
original game. We start with the Bertrand stage. The capacities
$k_i$ and $k_j$ are assumed to be given, and will be optimized in
the next section describing the Cournot stage. Hereafter, we assume
that the user type $\alpha$ is uniformly distributed. This
assumption was also used in \cite{Walrand} and \cite{Shetty}. We
will show how the price dynamics changes with more general
distributions of user type $\alpha$ in the last of this section.

\subsection{Price War with Long Jumps}
We derive the optimal price of MNO $i$, which is summarized in the
following lemmas:

\vskip 10pt \noindent {\bf {Lemma 1}}: {\it In the case that the
MNO's price $p_i$ is lower than its competitor's price $p_j$, the
optimal solution $p_i^L$ is:
\begin{eqnarray}
p_i^L  = \left\{ \begin{array}{l}
 \frac{{1 }}{2}\,\,\,\,\,\,\,\,\,\,\,\,\,\,\,\,\,\,\,{\text{if}}\,\,\,p_j  > \frac{{1 }}{2}\\
 p_j-\varepsilon\,\,\,\,\,\,\,{\text{if}}\,\,\,p_j  \le \frac{{1
}}{2}
 \end{array} \right.,
\end{eqnarray}}
{\it where $\varepsilon$ is a minimum unit of price level changes
and very small positive value.}

\vskip 10pt \noindent {\bf {Proof}}: Under the assumption that
$\alpha$ is uniformly distributed, we get the following equation
from the constraint of the optimization problem for the lower price
case:
\begin{eqnarray} \label{eq:demand_lower} d_i  = M\int_{p_i }^{1 - \frac{{d_i }}
{k_i M}} {f \left( \alpha  \right)d\alpha  } = M\left( {1 -
\frac{{d_i }} {k_i M} - p_i } \right).
\end{eqnarray}
\noindent We calculate $d_i$ and the objective function $p_id_i$
from Equation (\ref{eq:demand_lower}) as follows:
\begin{eqnarray} \label{eq:closed_demand_lower}
d_i  = \frac{{ k_i \left( {1 - p_i } \right)}} {k_i + 1}M, \,\,\,
p_i d_i = \frac{{k_i p_i \left( {1 - p_i } \right)}} {k_i + 1}M.
\end{eqnarray}
\noindent The objective function is a quadratic function whose
maximum is at $p_i=1/2$. Therefore, if $p_j>1/2$, then the optimal
solution will be $p_i^L=1/2$. Otherwise, the optimal solution will
be $p_i^L=p_j-\varepsilon$. \hfill $\blacksquare$

\vskip 10pt \noindent {\bf {Lemma 2}}: {\it In the case that the
MNO's price $p_i$ is higher than its competitor's price $p_j$, the
optimal solution $p_i^H$ is:
\begin{eqnarray}
p_i^H  = \left\{ \begin{array}{l}
 \frac{{k_j  + p_j }}{{k_j  + 1}}\,\,\,\,\,\,\,{\text{if}}\,\,\,p_j  \ge \frac{{1 - k_j
 }}{2}\\
 \frac{1}{2}\,\,\,\,\,\,\,\,\,\,\,\,\,\,\,\,\,{\text{if}}\,\,\,p_j  < \frac{{1 - k_j }}{2}
 \end{array} \right..
\end{eqnarray}}

\vskip 10pt \noindent {\bf {Proof}}: Under the assumption that
$\alpha$ is uniformly distributed, we get the following equation
from the constraint of the optimization problem for the higher price
case:
\begin{eqnarray} \label{eq:demand_higher}
d_i &=&M\int_{\max \left\{ {p_i ,1 - \frac{{d_j }} {k_j M}} \right\}
}^{1 - \frac{{d_i }} {k_i M}} {f \left( \alpha \right)d\alpha  }
\nonumber \\&=& M\left( {1 - \frac{{d_i }} {k_i M} - \max \left\{
{p_i ,1 - \frac{{d_j }} {k_j M}} \right\}} \right).
\end{eqnarray}
\noindent We calculate $d_i$ from Equation (\ref{eq:demand_higher})
as follows:
\begin{eqnarray} \label{eq:closed_demand_higher}
d_i  = \min \left\{{\frac{{ k_i \left(  {1 - p_i } \right)}}
{k_i+1}M,\frac{{k_i d_j }} {\left( {k_i + 1} \right) k_j}} \right\}.
\end{eqnarray}
Then, the objective function is: \setlength\arraycolsep{0.2pt}
\begin{eqnarray} \label{eq:objective_higher} p_i d_i  &=&  \min \left\{ \frac{{ k_i p_i \left(  {1 - p_i } \right)}}
{k_i+1} M, \frac{{k_i p_i d_j }} {\left( {k_i + 1} \right) k_j}
\right\} \nonumber \\ &=& \min \left\{ \frac{{ k_i p_i \left(  {1 -
p_i } \right)}} {k_i+1} M, \frac{{k_i p_i \left(  {1 - p_j } \right)
}} {\left( {k_i + 1} \right) \left( {k_j+ 1} \right)}M \right\}.
\end{eqnarray}
\noindent The second equality in Equation
(\ref{eq:objective_higher}) holds by Equation
(\ref{eq:closed_demand_lower}). In the minimum operator of the
objective function, the left side is a quadratic function whose
maximum is at $p_i=1/2$, and the right side is a linear function
whose slope is $k_i(1-p_j)M/((k_i+1)(k_j+1))$. Therefore, if
$p_j<(1-k_j)/2$, then the optimal solution is $p_i^H=1/2$, which is
the apex of the quadratic function. Otherwise, the optimal solution
is $p_i^H=(k_j+p_j)/(k_j+1)$, which is the intersection of the
quadratic and linear functions. \hfill $\blacksquare$

\vskip 10pt From Lemmas 1 and 2, we derive the MNO's best response
function (optimal strategy) in the duopoly competition.

\vskip 10pt \noindent {\bf Lemma 3}: {\it Given the competitor's
price $p_j$, the MNO's best response function $p_i^*$ is
\begin{itemize}
\item {\it Case 1} $\left( k_j < 1 \right)$:
\begin{eqnarray}
p_i^*  = \left\{ \begin{array}{l}
 \frac{1}{2}\,\,\,\,\,\,\,\,\,\,\,\,\,\,\,\,\,\,{\text{if}}\,\,\,p_j   >
 \frac{1}{2} \,\,\,{\text{or}}\,\,\, 0 \le p_j  < \frac{{1 - k_j }}{2}\\
 p_j  - \varepsilon \,\,\,\,\,\,{\text{if}}\,\,\,\frac{1}{{k_j  + 2}} < p_j  \le \frac{1}{2} \\
 \frac{{k_j  + p_j }}{{k_j  + 1}}\,\,\,\,\,\,\,\,{\text{if}}\,\,\,\frac{{1 - k_j }}{2} \le p_j  \le \frac{1}{{k_j  + 2}}
 \end{array} \right..
\end{eqnarray}
\end{itemize}

\begin{itemize}
\item {\it Case 2} $\left( k_j \ge 1 \right)$:
\begin{eqnarray}
p_i^*  = \left\{ \begin{array}{l}
 \frac{1}{2}\,\,\,\,\,\,\,\,\,\,\,\,\,\,\,\,\,\,{\text{if}}\,\,\,p_j  > \frac{1}{2} \\
 p_j  - \varepsilon \,\,\,\,\,\,{\text{if}}\,\,\,\frac{1}{{k_j  + 2}} < p_j  \le \frac{1}{2} \\
 \frac{{k_j  + p_j }}{{k_j  + 1}}\,\,\,\,\,\,\,\,{\text{if}}\,\,\,0 \le p_j  \le \frac{1}{{k_j  + 2}} \\
 \end{array} \right..
\end{eqnarray}
\end{itemize}}

\vskip 10pt \noindent {\bf {Proof}}: Using the results of Lemmas 1
and 2, we calculate the optimal values for the higher and lower
price cases as follows:
\begin{eqnarray}
p_i^L d_i^L  = \left\{ \begin{array}{l}
 \frac{{k_i}} {4 \left( k_i + 1 \right)}M
 \,\,\,\,\,\,\,\,\,\,\,\,\,\,\,\,\,\,\,\,\
 \,\,\,\,\,\,\,\,\,\,\,\,\,\,\,\,\,\,\,\,\,\,\,\,\,\,\,\,\,\,\,\,\,\,\,\,\,\,\,\,\,\,\,{\text{if}}\,\,\,p_j  > \frac{{1 }}{2}\\
 \frac{{k_i \left( p_j - \varepsilon \right) \left( {1 - p_j + \varepsilon }
 \right)}} {k_i + 1}M \approx \frac{{k_i p_j  \left( {1 - p_j  } \right)}} {k_i + 1}M
\,\,\,{\text{if}}\,\,\,p_j  \le \frac{{1 }}{2}
 \end{array} \right., \nonumber
\end{eqnarray}

\begin{eqnarray}
p_i^H d_i^H  = \left\{ \begin{array}{l} \frac{{k_i}} {4 \left( k_i +
1 \right)}M
\,\,\,\,\,\,\,\,\,\,\,\,\,\,\,\,\,\,\,\,\,\,\,\,\,\,{\text{if}}\,\,\,p_j
\ge \frac{{1 - k_j
 }}{2}\\
 \frac{{k_i \left(  {k_j + p_j } \right) \left(  {1 - p_j } \right) }}
{\left( {k_i + 1} \right) \left( {k_j+ 1} \right)^2}M
\,\,\,\,\,\,\,{\text{if}}\,\,\,p_j  < \frac{{1 - k_j }}{2}
 \end{array} \right.. \nonumber
\end{eqnarray}

\noindent Above all, we consider Case 1 ($k_j<1$). In this case, we
calculate the best response function of MNO $i$ as follows:
\begin{itemize}
\item {\it If $p_j>1/2$}: To compare the optimal values of the higher and lower
price cases, we calculate the following: \setlength\arraycolsep{1pt}
\begin{eqnarray} p_i^L d_i^L - p_i^H d_i^H &=& \frac{{k_i}M} {4
\left( k_i + 1 \right)} - \frac{{k_i \left(  {k_j + p_j } \right)
\left(  {1 - p_j } \right) }M} {\left( {k_i + 1} \right) \left(
{k_j+ 1} \right)^2}
 \nonumber \\ &=& \frac{k_i \left( k_i+2p_j-1 \right)^2M}{4 \left( k_i+1 \right) \left( k_j+1
\right)^2}. \nonumber
\end{eqnarray}
This value is positive because we consider the duopoly market (i.e.,
$k_i$ and $k_j$ are positive) and assume $p_j>1/2$. Therefore, the
best response function is $p_i^*=p_i^L=1/2$.
\end{itemize}

\begin{itemize}
\item {\it If $(1-k_j)/2 \le p_j \le 1/2$}: To compare the optimal values of the higher and lower
price cases, we calculate the following:
\setlength\arraycolsep{0.3pt}
\begin{eqnarray}
&&p_i^L d_i^L - p_i^H d_i^H \nonumber \\ &&= \frac{{k_i p_j  \left(
{1 - p_j } \right)}M} {k_i + 1} - \frac{{k_i \left(  {k_j + p_j }
\right) \left( {1 - p_j } \right) }M} {\left( {k_i + 1} \right)
\left( {k_j+ 1} \right)^2} \nonumber \\ &&= \frac{k_i k_j \left( k_j
+ 2 \right) \left( 1 - p_j \right) M}{\left( k_i+1 \right) \left(
k_j+1 \right)^2} \left( p_j - \frac{1}{k_j+2}  \right). \nonumber
\end{eqnarray}
All values except $p_j-1/(k_j+2)$ are positive because we consider
the duopoly market (i.e., $k_i$ and $k_j$ are positive) and assume
$(1-k_j)/2 \le p_j \le 1/2$. Therefore, if $p_j > 1/(k_j+2)$, then
the best response function is $p_i^*=p_i^L=p_j-\varepsilon$.
Otherwise, the best response function is
$p_i^*=p_i^H=(k_j+p_j)/(k_j+1)$.
\end{itemize}

\begin{itemize}
\item {\it If $0 \le p_j < (1-k_j)/2$}: To compare the optimal values of the higher and lower
price cases, we calculate the following: \setlength\arraycolsep{1pt}
\begin{eqnarray}
p_i^H d_i^H - p_i^L d_i^L &=& \frac{{k_i}M} {4 \left( k_i + 1
\right)} - \frac{{k_i p_j  \left( {1 - p_j  } \right)}M} {k_i + 1}
\nonumber
\\ &=& \frac{k_iM}{k_i + 1} \left( \frac{1}{4} - p_j\left(1-p_j\right)
\right). \nonumber
\end{eqnarray}
This value is positive under the assumption $0 \le p_j < (1-k_j)/2$.
Therefore, the best response function is $p_i^*=p_i^H=1/2$.
\end{itemize}

\noindent Noting that the intervals $[0,(1-k_j)/2)$ and
$[(1-k_j)/2,1/(k_j+2)]$ in Case 1 are merged into an interval
$[0,1/(k_j+2)]$ in Case 2, we can derive the MNO's best response
function in Case 2 similarly. \hfill $\blacksquare$

\vskip 10pt Using Lemma 3, we describe two MNOs dynamics below.

\vskip 10pt \noindent {\bf Proposition 1}: {\it In the duopoly
competition of two MNOs, there is no pure Nash equilibrium, and
price levels increase and decrease periodically.}\footnote{Even
though there is no Nash equilibrium in pure strategies, there can be
a Nash equilibrium in mixed strategies. This means that the MNOs can
randomize their service prices. However, in the sequential game that
each MNO can change its price after observing a competitor's, the
best response function-based strategy will be used rather than the
random strategy. This is the reason that we focus on the price
dynamics.}

\vskip 10pt \noindent {\bf {Proof}}: Suppose for a contradiction
that a pure Nash equilibrium point $(p_i^{NE},p_j^{NE})$ exists. We
first consider Case 1 ($k_j<1$) as follows:
\begin{itemize}
\item If $p_j^{NE}>1/2$, then $p_i^{NE}$ should be
$p_i^{NE}=1/2$. Then, $p_j^{NE}$ should be
$p_j^{NE}=1/2-\varepsilon$, which contradicts the assumption that
$p_j^{NE}>1/2$.
\end{itemize}

\begin{itemize}
\item If $1/(k_j+2) < p_j^{NE} \le 1/2$, then $p_i^{NE}$ should be
$p_i^{NE}=p_j^{NE}-\varepsilon < 1/2$. Then, $p_j^{NE}$ should
satisfy at least one of the following equations:
\begin{eqnarray}
p_j^{NE}&=&p_i^{NE}-\varepsilon = p_j^{NE}-2\varepsilon,  \label{eq:proposition1_eq1} \\
p_j^{NE}&=&\frac{k_i+p_i^{NE}}{k_i+1}=\frac{k_i+p_j^{NE}-\varepsilon}{k_i+1}.
\label{eq:proposition1_eq2}
\end{eqnarray}
Obviously, Equation (\ref{eq:proposition1_eq1}) has a contradiction.
From Equation (\ref{eq:proposition1_eq2}), we find that $p_j^{NE}$
is $p_j^{NE}=1-\varepsilon/k_i \approx 1$, which contradicts the
assumption that $p_j^{NE} \le 1/2$.
\end{itemize}

\begin{itemize}
\item If $(1-k_j)/2 \le p_j^{NE} \le 1/(k_j+2)$, then $p_i^{NE}$ should be
$p_i^{NE}=(k_j+p_j^{NE})/(k_j+1)$. Then, $p_j^{NE}$ should satisfy
at least one of the following equations:
\begin{eqnarray}
p_j^{NE}&=&\frac{1}{2},  \label{eq:proposition1_eq3} \\
p_j^{NE}&=&p_i^{NE}-\varepsilon=\frac{k_j+p_j^{NE}}{k_j+1}-\varepsilon, \label{eq:proposition1_eq4}\\
p_j^{NE}&=&\frac{k_i+p_i^{NE}}{k_i+1}=\frac{k_i+\frac{k_j+p_j^{NE}}{k_j+1}}{k_i+1}.
\label{eq:proposition1_eq5}
\end{eqnarray}
Based on the assumption that $p_j^{NE} \le 1/(k_j+2)$, Equation
(\ref{eq:proposition1_eq3}) means that $k_j=0$ and
$p_i^{NE}=p_j^{NE}=1/2$. This has a contradiction because
$p_i^{NE}=1/2$ and $p_j^{NE}=1/2$ are not the best response to each
other. Using Equation (\ref{eq:proposition1_eq4}), we calculate that
$p_j^{NE}$ is $p_j^{NE}=1-\varepsilon-\varepsilon/k_j \approx 1$,
which contradicts the assumption that $p_j^{NE} \le 1/(k_j+2)$.
Using Equation (\ref{eq:proposition1_eq5}), we calculate that
$p_j^{NE}$ is $p_j^{NE}=1$, which contradicts the assumption that
$p_j^{NE} \le 1/(k_j+2)$.
\end{itemize}

\begin{itemize}
\item If $0 \le p_j^{NE} < (1-k_j)/2 $, then $p_i^{NE}$ should be
$p_i^{NE}=1/2$. This means that $p_j^{NE}$ should be
$p_j^{NE}=1/2-\varepsilon \approx 1/2$, which contradicts the
assumption that $p_j^{NE} < (1-k_j)/2$.
\end{itemize}

\noindent Therefore, we conclude that there is no pure Nash
equilibrium point in Case 1. The only difference between Case 1 and
Case 2 ($k_j \ge 1$) is that the intervals $[0,(1-k_j)/2)$ and
$[(1-k_j)/2,1/(k_j+2)]$ in Case 1 are merged in the interval
$[0,1/(k_j+2)]$ in Case 2. Therefore, we can use similar proof for
Case 2 and conclude that there is no pure Nash equilibrium point in
Case 2. \hfill $\blacksquare$

\begin{figure*}[t]
\centerline{\epsfig{figure=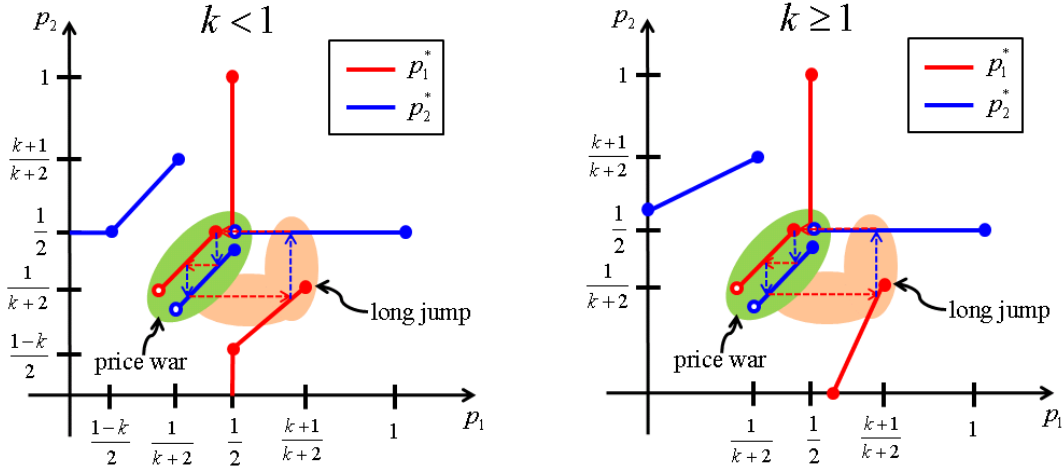,height=2.5in,clip=;}}
\caption{Best response functions of both MNOs. We assume that the
MNOs' capacities are equal ($k_1=k_2=k$).} \label{BR}
\end{figure*}

\begin{figure*}[t]
\centerline{\epsfig{figure=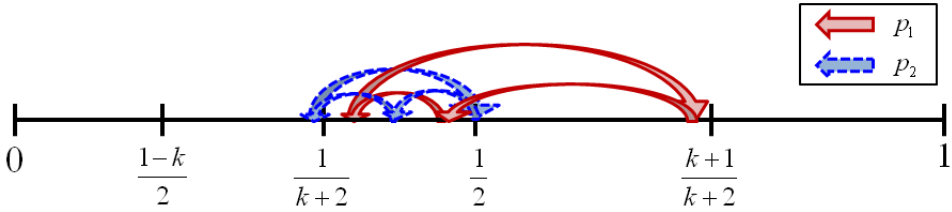,height=1.1in,clip=;}}
\caption{Price war with long jumps ($k_1=k_2=k$).} \label{long_jump}
\end{figure*}

\vskip 10pt Figure \ref{BR} shows the MNOs' best response functions.
In the figure, we assume the symmetric capacity ($k_1=k_2=k$). We
see how each MNO sequentially decreases its price to a little lower
level than its competitor when the competitor's price is within
$(1/(k+2),1/2]$. This is a {\it price war} \cite{Chiu, Tan}. After
that, if the competitor's price is less than or equal to $1/(k+2)$,
then one MNO will increase its price to $(k+1)/(k+2)$. This is a
{\it long jump}. Then, the competitor sets its price to $1/2$. This
situation is repeated periodically and there is no equilibrium
point. We call this {\it price war with long jumps}. We plot Figure
\ref{long_jump} to illustrate the process of the price war with long
jumps. In the real world, this kind of price dynamics tends to occur
by change of the subsidy amount (see Figure \ref{subsidy}).

\subsection{Regulation for Convergence}
The price war with long jumps is not desirable because of
instability and inefficiency. We suggest a simple regulation that
leads to an equilibrium point of price levels, which is
Pareto-optimal. More details are contained in Lemma 4 and
Proposition 2.

\vskip 10pt \noindent {\bf Lemma 4}: {\it A regulation that limits
the number of price level changes makes the price levels converge to
an equilibrium point $(p^E_i,p^E_j)$ as follows:
\begin{eqnarray}
(p^E_i,p^E_j)  = \left\{ \begin{array}{l} \left( {\frac{1}{{k_i  +
2}},\frac{{k_i  + 1}}{{k_i  + 2}}} \right)
 \,\,\,\,\,\,\,\,\,\,\,\,\,\,\,\,\,\,\,\,\,\,\,\,\,\,\,\,\,\,\,\,\,\,\,\,\,\,\,\,\,\,\,\,\,\,{\text{if}}\,\,\, k_i < 2k_j\\\left(
{\frac{1}{{k_i  + 2}},\frac{{k_i  + 1}}{{k_i  + 2}}} \right)\,\,
{\text{or}}\,\, \left( {\frac{{2k_j  + 1}}{{2\left( {k_j + 1}
\right)}},\frac{1}{2}} \right)\,\,\,{\text{if}}\,\,\, k_i = 2k_j\\
\left( {\frac{{2k_j  + 1}}{{2\left( {k_j  + 1}
\right)}},\frac{1}{2}} \right)
 \,\,\,\,\,\,\,\,\,\,\,\,\,\,\,\,\,\,\,\,\,\,\,\,\,\,\,\,\,\,\,\,\,\,\,\,\,\,\,\,\,\,\,\,\,\,\,\,{\text{if}}\,\,\, k_i >
2k_j
 \end{array} \right. \nonumber,
\end{eqnarray}}
{\it where the last opportunity for a price level change is given to
MNO $j$.}

\vskip 10pt \noindent {\bf {Proof}}: When we use the regulation, two
MNOs dynamics is modeled as a finite sequential game and we can
calculate the equilibrium point using the backward induction method.
We first consider Case 1 ($k_j<1$). Let $t$ denote the last stage of
the price level change. Then, MNO $i$ will choose the best strategy
at the $(t-1)$ stage (i.e., its last choice stage) in order to
maximize its revenue. To calculate the best strategy for MNO $i$, we
divide the strategy set of MNO $i$ into four disjoint subsets. Let
$p_i^*(t-1)$ and $p_j^*(t)$ denote the optimal strategy of MNO $i$
at the $t-1$ stage and the optimal strategy of MNO $j$ at the last
stage in each subset. Using Lemma 3, we calculate $p_i^*(t-1)$ in
each subset. Then, from Equations (\ref{eq:closed_demand_lower}) and
(\ref{eq:objective_higher}), we calculate the revenue of MNO $i$
$r_i^*(t)$ as follows:

\noindent {\it Strategy 1} $(p_i (t - 1)>1/2)$:
\setlength\arraycolsep{1pt}
\begin{eqnarray}
p_j^*(t) &=& \frac{1} {2} \, \xrightarrow{{}} \, p_i^*(t-1) =
\frac{k_j+\frac{1}{2}} {k_j+1}=\frac{2k_j+1} {2(k_j+1)},\nonumber \\
r_i^*(t)&=& \frac{{k_i p_i^*(t) \left( {1 - p_i^*(t) } \right)}}
{k_i + 1}M=\frac{k_i \left(2k_j + 1 \right)} {4 \left(k_i + 1
\right)\left(k_j + 1 \right)^2}M. \nonumber
\end{eqnarray}

\noindent {\it Strategy 2} $(1/(k_i+2) < p_i (t - 1) \le 1/2)$:
\setlength\arraycolsep{1pt}
\begin{eqnarray}
p_j^*(t) &=& p_i(t-1)  - \varepsilon
\, \xrightarrow{{}} \,p_i^*(t-1) = \frac{1} {2},\nonumber \\
r_i^*(t)&=&\frac{{k_i p_i^*(t-1) \left(  {1 - p_j^*(t) } \right) }}
{\left( {k_i + 1} \right) \left( {k_j+ 1} \right)}M=\frac{{k_i
 \left(  1+2\varepsilon \right) }} {4\left( {k_i + 1}
\right) \left( {k_j+ 1} \right)}M. \nonumber
\end{eqnarray}

\noindent {\it Strategy 3} $((1-k_i)/2 \le p_i (t - 1) \le
1/(k_i+2))$: \setlength\arraycolsep{1pt}
\begin{eqnarray}
p_j^*(t) &=& \frac{k_i+p_i(t-1)}
{k_i+1}\, \xrightarrow{{}} \,p_i^*(t-1) =\frac{1} {k_i+2},\nonumber \\
r_i^*(t)&=&\frac{{k_i p_i^*(t) \left( {1 - p_i^*(t) } \right)}} {k_i
+ 1}M=\frac{k_i} {\left(k_i + 2\right)^2}M. \nonumber
\end{eqnarray}

\noindent {\it Strategy 4} $(0 \le p_i (t - 1) < (1-k_i)/2 )$:
\setlength\arraycolsep{1pt}
\begin{eqnarray}
p_j^*(t) &=& \frac{1} {2}\, \xrightarrow{{}} \,p_i^*(t-1) =
\frac{1-k_i} {2} - \varepsilon,\nonumber \\
r_i^*(t)&=&\frac{{k_i p_i^*(t) \left( {1 - p_i^*(t) } \right)}} {k_i
+ 1}M=\frac{k_i \left(1-\left( k_i +\varepsilon \right)^2 \right)}
{4 \left(k_i + 1 \right)}M. \nonumber
\end{eqnarray}

\noindent From these equations, we prove that some strategies are
strictly dominated as follows:
\begin{itemize}
\item Strategy 2 is strictly dominated by Strategy 1:
\begin{eqnarray}
\frac{k_i \left(2k_j + 1 \right)} {4 \left(k_i + 1 \right)\left(k_j
+ 1 \right)^2}M - \frac{{k_i
 \left(  1+2\varepsilon \right) }} {4\left( {k_i + 1}
\right) \left( {k_j+ 1} \right)}M \nonumber \\=\frac{k_i} {4\left(
{k_i + 1} \right) \left( {k_j+ 1} \right)} \left( \frac{k_j}{k_j +
1} -2 \varepsilon \right)M>0. \nonumber
\end{eqnarray}
\end{itemize}

\begin{itemize}
\item Strategy 4 is strictly dominated by Strategy 3:
\begin{eqnarray}
&&\frac{k_i} {\left(k_i + 2\right)^2}M-\frac{k_i \left(1-\left( k_
+\varepsilon \right)^2 \right)} {4 \left(k_i + 1 \right)}M \nonumber
\\ &&>\frac{k_i} {\left(k_i + 2\right)^2}M-\frac{k_i \left(1-k_i^2
\right)} {4 \left(k_i + 1 \right)}M \nonumber \\ &&=\frac{k_i^3
\left(k_i+3 \right)}{4 \left( k_i+2 \right)^2}M>0. \nonumber
\end{eqnarray}
\end{itemize}

\noindent Therefore, the MNO $i$ never chooses Strategies 2 and 4.
To finish this proof, we compare Strategies 1 and 3 as follows:
\begin{eqnarray}
\frac{k_i \left(2k_j + 1 \right)} {4 \left(k_i + 1 \right)\left(k_j
+ 1 \right)^2}M - \frac{k_i} {\left(k_i + 2\right)^2}M  \nonumber \\
=\frac{k_i \left(k_i + 2\left(k_i + 1 \right)k_j \right)
\left(k_i-2k_j \right)}{4\left(k_i + 1 \right)\left(k_i + 2
\right)^2\left(k_j + 1 \right)^2}M. \nonumber
\end{eqnarray}
If $k_i<2k_j$, then MNO $i$ will set its price to $1/(k_i+2)$ and
MNO $j$ will set its price to $(k_i+1)/(k_i+2)$. Therefore, in this
case, the equilibrium point is
$(p_i^E,p_j^E)=(1/(k_i+2),(k_i+1)/(k_i+2))$. Likewise, if
$k_i>2k_j$, then the equilibrium point is
$(p_i^E,p_j^E)=((2k_j+1)/(2(k_j+1)),1/2)$. Note that, if $k_i=2k_j$,
then both equilibriums are possible. The only difference between
Case 1 and Case 2 is that the intervals $[0,(1-k_j)/2)$ and
$[(1-k_j)/2,1/(k_j+2)]$ in Case 1 are merged into the interval
$[0,1/(k_j+2)]$ in Case 2. Therefore, we can use a similar proof for
Case 2 and conclude that the equilibrium in Case 2 is same to that
in Case 1. \hfill $\blacksquare$

\vskip 10pt To verify the efficiency of the regulation, we need to
check the Pareto-optimality of the equilibrium point. A sufficient
condition for Pareto-optimality is given in the following lemma:

\vskip 10pt \noindent {\bf {Lemma 5}}: {\it Let subscripts $l$ and
$h$ denote MNOs whose prices are lower and higher than that of their
competitor, respectively. If $p_l \le 1/2$, $p_h \ge 1/2$ and $p_h
\ge (k_l+p_l)/(k_l+1)$, then $(p_l,p_h)$ and $(p_h,p_l)$ are
Pareto-optimal.}

\vskip 10pt \noindent {\bf {Proof}}: If $p_l<1/2$, then MNO $l$ can
increase its revenue by increasing its price toward $1/2$. However,
it always makes MNO $h$'s revenue decrease. If $p_l=1/2$, then MNO
$l$ cannot increase its revenue. Therefore, MNO $l$ cannot increase
its revenue without decreasing the revenue of MNO $h$. Likewise, MNO
$h$ cannot increase its revenue without decreasing the revenue of
MNO $l$ because the only method to increase the revenue of MNO $h$
in the perfectly segmented condition ($p_h \ge (k_l+p_l)/(k_l+1)$)
is to decrease $p_l$ and $p_h$ simultaneously. Therefore, the point
that satisfies the conditions in Lemma 5 is Pareto-optimal. \hfill
$\blacksquare$

\vskip 10pt Lemma 5 indicates that the MNOs' prices should be
sufficiently separated (i.e., perfectly segmented market) for
Pareto-optimality. The equilibrium point under our suggested
regulation (Lemma 4) satisfies the conditions in Lemma 5. Therefore,
the equilibrium point is Pareto-optimal. This finding is summarized
below.

\vskip 10pt \noindent {\bf Proposition 2}: {\it A regulation that
limits the number of price level changes makes the price levels
converge to an equilibrium point that is Pareto-optimal.}

\vskip 10pt \noindent The price war with long jumps occurs due to
the MNOs' short-sighted way of thinking. However, the MNOs cannot
use a myopic strategy under Proposition 2.

We can find similar regulations of Proposition 2. In South Korea,
the market-dominating enterprise (SKT) cannot change its service
price without government permission. This is similar to giving the
other MNOs (KT and LGU+) the last opportunity for a price level
change. The price regulation also can be implemented by restricting
subsidization. In other words, notifying MNOs that subsidization
will be banned after a certain time is equal to the regulation that
limits the number of price level changes. For instance, Korea
Communications Commission (KCC) actually prohibited subsidization by
MNOs from 2003 to 2008. This kind of subsidy regulation also can be
observed in Finland \cite{Tallberg}. Another way that brings the
same effect of such price regulation is to regulate the time
interval between price level changes. If MNOs cannot change their
service price or subsidy levels for a long time, then they will use
far-sighted strategies, which leads to the equilibrium point
described in Lemma 4.

\begin{figure}[t]
\centerline{\epsfig{figure=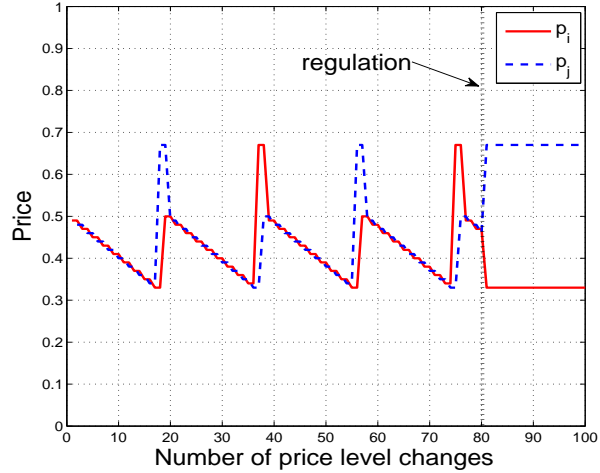,width=3.5in,
height=2.7in, clip=;}} \caption{Price level changes of both MNOs
(uniform user type case).} \label{price_war_uniform}
\end{figure}

\begin{figure*}
\centerline{
\epsfig{figure=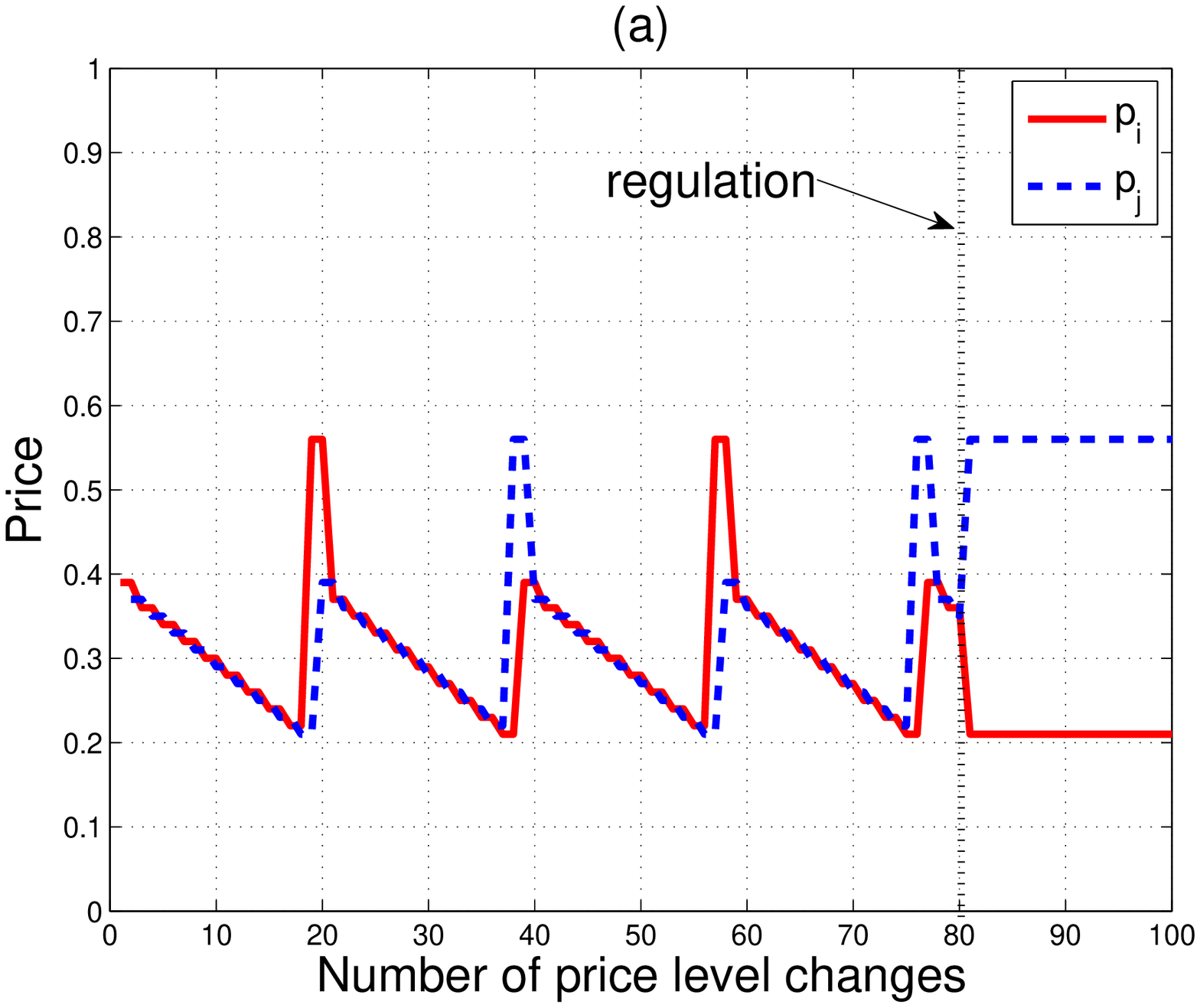,height=1.7in,clip=;}
\epsfig{figure=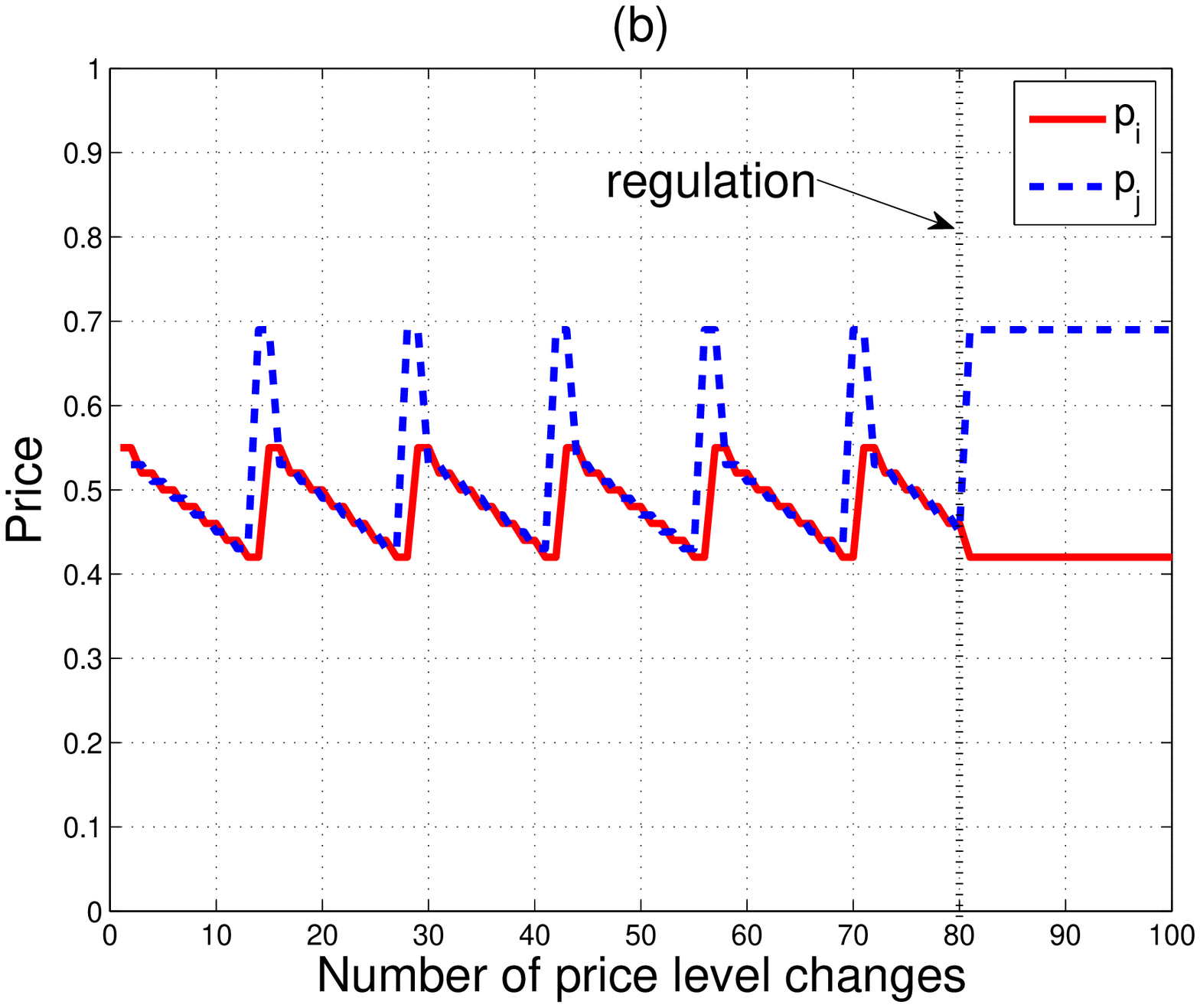,height=1.7in,clip=;}
\epsfig{figure=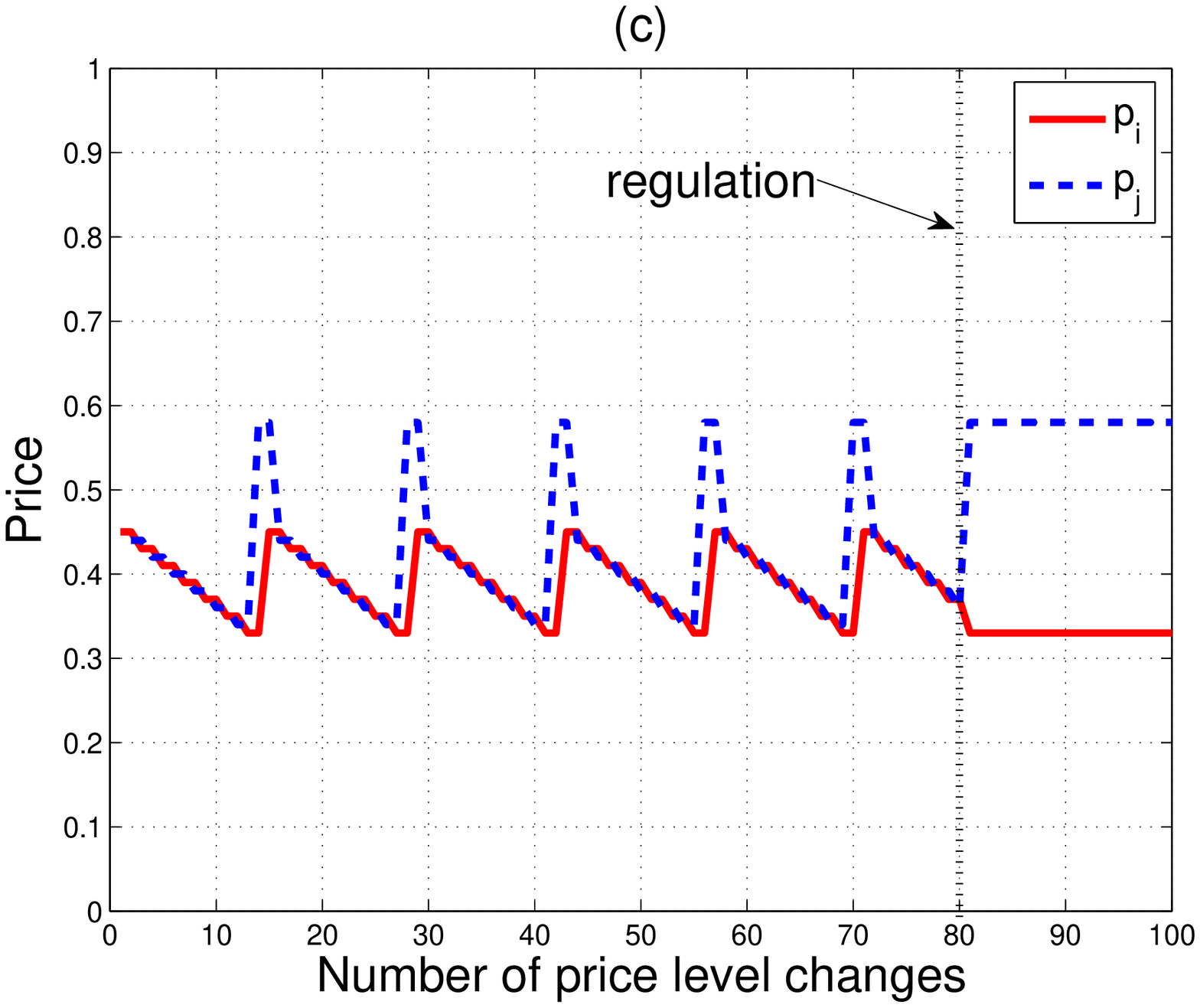,height=1.7in,clip=;}}
\caption{Price level changes of both MNOs in non-uniform user type
cases: (a) $f_1(\alpha)$. (b) $f_2(\alpha)$. (c) $f_3(\alpha)$.}
\label{price_war_nonuniform}
\end{figure*}

To verify our analysis, we conduct simulations, which show the price
level changes of both MNOs. In the simulations, the minimum unit of
price level changes and each MNO's network capacity are set to
$\varepsilon=0.01$ and $k_i=k_j=1$. Also, we set each MNO's initial
price to $0.01$ and apply the price regulation that limits the
number of price level changes to $80$. Figure
\ref{price_war_uniform} illustrates the results, where both MNOs
initially decrease their prices repeatedly but one MNO suddenly
increases its price when the competitor's price is lower than some
threshold (i.e., price war with long jumps). After the price
regulation, the prices converge on an equilibrium point, which
coincides with our analysis.

\subsection{Non-uniform User Type Case}
So far, we assume that user type $\alpha$ of (\ref{eq:demand}) is
uniformly distributed. We will now see how the price dynamics
changes with more general distributions of $\alpha$ by means of
simulations. For this, we adopt three additional distributions of
$\alpha$ in \cite{Lee} as follows:
\begin{eqnarray}
f_1(\alpha)&=&2-2\alpha, \nonumber \\ f_2(\alpha)&=&2\alpha, \nonumber \\
f_3 \left( \alpha \right) &=& \left\{ \begin{array}{l}
 4\alpha, \,\,\,\,\,\,\,\,\,\,\,\,\,\,\, {\rm{if}} \,\, 0 \le \alpha  \le \frac{1}{2}, \\
 4 - 4\alpha, \,\,\,\,\, {\rm{if}}\,\, \frac{1}{2} < \alpha  \le 1. \\
 \end{array} \right.  \nonumber
\end{eqnarray}
Using these distributions, we can reflect network scenarios
consisting of high population of users having low, high and middle
user type, respectively.

Figure \ref{price_war_nonuniform} shows the results, where price war
with long jumps occurs like the uniform user type case. Moreover, in
the non-uniform user type cases, the prices always converge on an
equilibrium point after the price regulation. The equilibrium price
tends to be biased towards the high density of user type.

\section{Cournot Stage: Capacity Competition}
In this section, we combine the result of the Bertrand stage with
that of the Cournot stage. That is, using the results of Section IV,
we rewrite the optimization problem (Equation
(\ref{eq:cournot_opt_problem})) of the Cournot stage and solve it.
The main motivation of this section is to completely understand the
competitive actions of each MNO, and thus to derive the optimal
response of the regulator.

\subsection{Characteristics of Communications Service Duopoly Market}
Without loss of generality, we assume that the last opportunity for
a price level change is given to MNO $j$. Using the revenue
equations (Equations (\ref{eq:closed_demand_lower}) and
(\ref{eq:objective_higher})), we calculate the revenue function
$f_i^R \left( {k_i ,k_j } \right)$ at the equilibrium price of the
Bertrand stage (Lemma 4). We use the linearly increasing cost
function as in \cite{Phillips}. That is, $f_i^C \left( {k_i }
\right)= \gamma M k_i$, where $\gamma$ is the unit cost per
capacity. Then, we can rewrite the optimization problem (Equation
(\ref{eq:cournot_opt_problem})) of the Cournot stage, which is
divided into two cases.
\begin{itemize}
\item {\it Case 1} $\left( k_i \le 2k_j  \right)$:
\begin{eqnarray}
 \mathop {\max }\limits_{k_i \ge 0} &&\,\,\,\, \frac{{Mk_i }}{{\left( {k_i  + 2} \right)^2 }} - \gamma Mk_i,  \\
 \mathop {\max }\limits_{k_j \ge 0} &&\,\,\,\,  \frac{{M\left( {k_i  + 1} \right)}}{{\left( {k_i  + 2} \right)^2 }}\frac{{k_j }}{{k_j  + 1}} - \gamma
 Mk_j.
\end{eqnarray}
\end{itemize}

\begin{itemize}
\item {\it Case 2} $\left( k_i \ge 2k_j \right)$:
\begin{eqnarray}
 \mathop {\max }\limits_{k_i \ge 0} &&\,\,\,\, \frac{{M\left( {2k_j  + 1} \right)}}{{4\left( {k_j  + 1} \right)^2 }}\frac{{k_i }}{{k_i  + 1}} - \gamma Mk_i,  \\
 \mathop {\max }\limits_{k_j \ge 0} &&\,\,\,\, \frac{M}{4}\frac{{k_j }}{{k_j  + 1}} - \gamma
 Mk_j.
\end{eqnarray}
\end{itemize}

To solve these optimization problems, we need some mathematical
knowledge given in the lemmas below.

\vskip 10pt \noindent {\bf Lemma 6}: {\it Consider the following
optimization problem:
\begin{eqnarray}
\mathop {\max }\limits_{x \ge 0} \,\,\,\,\, \frac{{bx}}{{x + a}} -
cx,
\end{eqnarray}
where $a$, $b$ and $c$ are positive. Then, the optimal solution
$x^*$ is
\begin{eqnarray}
x^*  =\max \left\{ {0,\sqrt {\frac{{ab}}{c}}  - a} \right\}.
\end{eqnarray}}

\vskip 10pt \noindent {\bf {Proof}}: We calculate the first and
second order derivatives of the objective function as follows:
\begin{eqnarray}
\left( {\frac{{bx}}{{x + a}} - cx} \right)^\prime   &=&
\frac{{ab}}{{\left( {x + a} \right)^2 }} - c, \\ \left(
{\frac{{bx}}{{x + a}} - cx} \right)^{\prime \prime }  &=& \frac{{ -
2ab\left( {x + a} \right)}}{{\left( {x + a} \right)^4 }}.
\end{eqnarray}
The problem is a convex optimization problem because the second
order derivative is always negative in the feasible set. Therefore,
using the first order condition, we calculate the optimal solution
as follows:
\begin{eqnarray}
x^*  =\max \left\{ {0,\sqrt {\frac{{ab}}{c}}  - a} \right\}.
\end{eqnarray} \hfill $\blacksquare$

\vskip 10pt \noindent {\bf Lemma 7}: {\it Consider the following
optimization problem:
\begin{eqnarray}
\mathop {\max }\limits_{x \ge 0} \,\,\,\,\, \frac{{bx}}{ {\left( {x
+ a} \right)^2 }} - cx,
\end{eqnarray}
where $a$, $b$ and $c$ are positive. Then, the optimal solution
$x^*$ is
\begin{eqnarray}
x^*  = \max \biggr\{ 0, - a + \sqrt[3]{{\frac{{ab}}{c} + \sqrt
{\frac{{a^2 b^2 }}{{c^2 }} + \frac{{b^3 }}{{27c^3 }}} }} \nonumber
\\+
\sqrt[3]{{\frac{{ab}}{c} - \sqrt {\frac{{a^2 b^2 }}{{c^2 }} +
\frac{{b^3 }}{{27c^3 }}} }} \biggr\}.
\end{eqnarray}}

\vskip 10pt \noindent {\bf {Proof}}: We calculate the first and
second order derivatives of the objective function as follows:
\begin{eqnarray}
\left( {\frac{{bx}}{{\left( {x + a} \right)^2 }} - cx}
\right)^\prime   &=& \frac{{ - bx + ab}}{{\left( {x + a} \right)^3
}} - c, \\ \left( {\frac{{bx}}{{\left( {x + a} \right)^2 }} - cx}
\right)^{\prime \prime }  &=& \frac{{2bx - 4ab}}{{\left( {x + a}
\right)^4 }}.
\end{eqnarray}
The objective function is partially concave because the second order
derivative is positive when $x>2a$. However, from the first order
derivative, we know that the objective function decreases as $x$
increases when $x>a$. Therefore, we only consider $0 \le x \le a$ as
the feasible set, and the optimization problem becomes a convex
optimization problem in the set. Then, using the first order
condition and the root formula of the third order equation, we
calculate the optimal solution $x^*$. \hfill $\blacksquare$

\vskip 10 pt Using Lemmas 6 and 7, we calculate the MNOs' optimal
solutions $k^*_i$ and $k^*_j$ as follows:
\begin{itemize}
\item {\it Case 1} $\left( k_i \le 2k_j  \right)$:
\begin{eqnarray}
k_i^*  &=& \max \biggr\{ 0, - 2 + \sqrt[3]{{\frac{2}{\gamma } +
\sqrt {\frac{4}{{\gamma ^2 }} + \frac{1}{{27\gamma ^3 }}} }}
\nonumber \\&+& \sqrt[3]{{\frac{2}{\gamma } - \sqrt
{\frac{4}{{\gamma ^2 }} +
\frac{1}{{27\gamma ^3 }}} }} \biggr\},\\
k_j^*  &=& \max \left\{ {0,\sqrt {\frac{{k_i  + 1}}{{\left( {k_i  +
2} \right)^2 \gamma }}}  - 1} \right\}.
\end{eqnarray}
\end{itemize}

\begin{itemize}
\item {\it Case 2} $\left( k_i \ge 2k_j \right)$:
\begin{eqnarray}
k_i^*  &=& \max \left\{ {0,\sqrt {\frac{{2k_j  + 1}}{{4\left( {k_j +
1} \right)^2 \gamma }}}  - 1} \right\},\\
k_j^*  &=& \max \left\{ {0,\sqrt {\frac{1}{{4\gamma }}}  - 1}.
\right\}
\end{eqnarray}
\end{itemize}
From this result, we find the equilibrium of the original two-stage
game as follows:

\begin{figure*}
\centerline{ \epsfig{figure=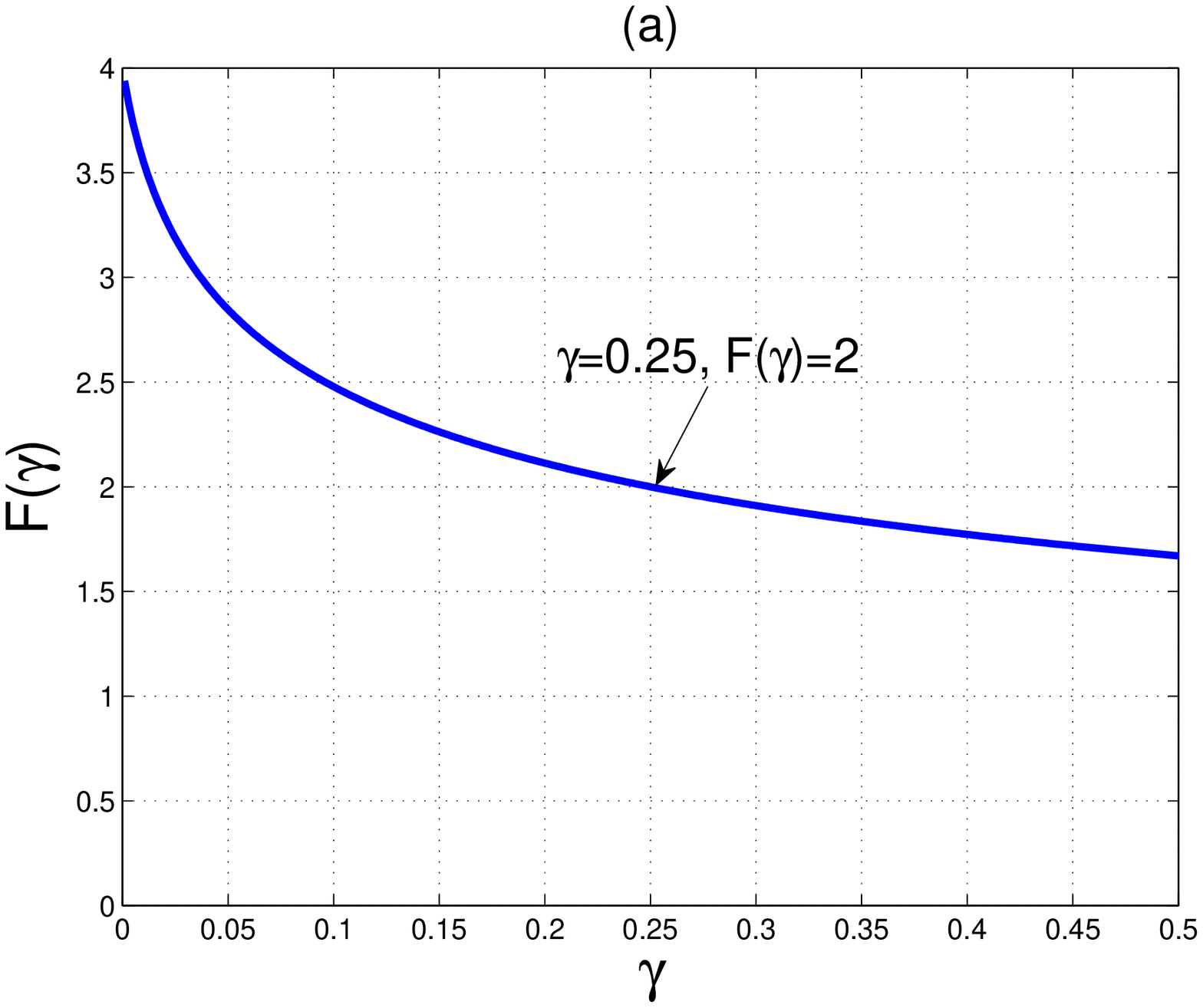,height=1.7in,clip=;}
\epsfig{figure=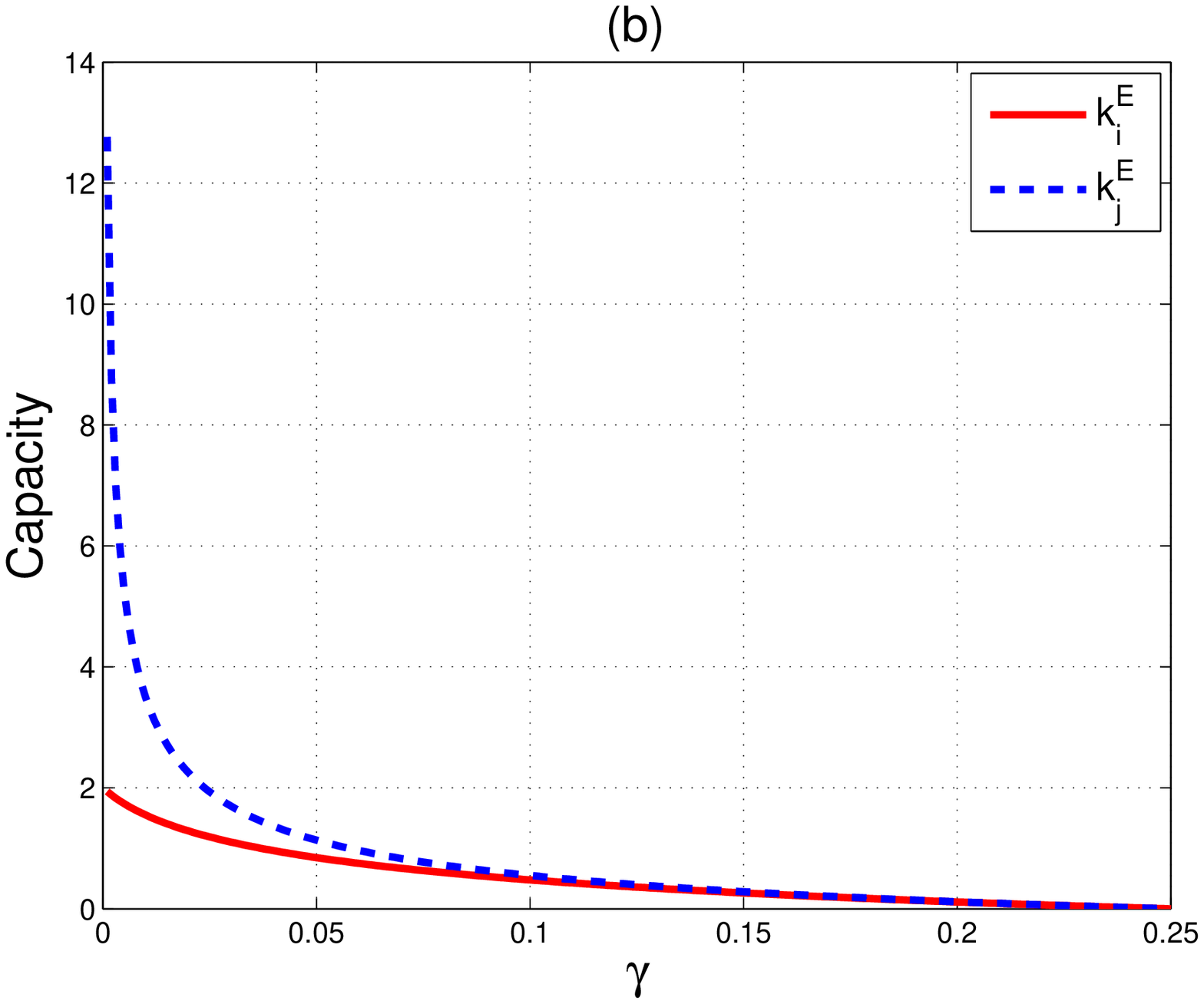,height=1.7in,clip=;}
\epsfig{figure=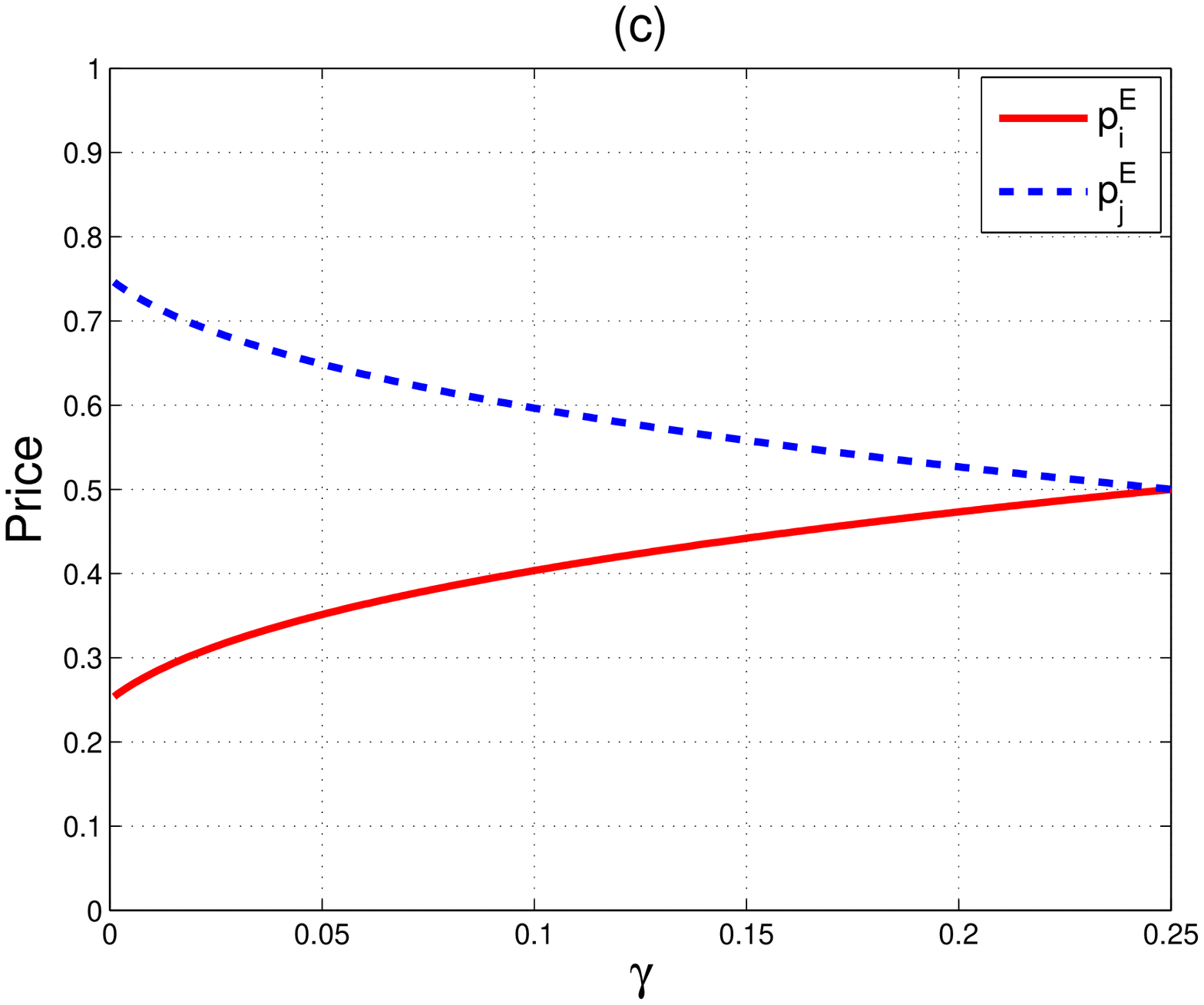,height=1.7in,clip=;}} \caption{(a)
Feasibility function $F(\gamma)$. (b) Equilibrium capacities $k^E_i$
and $k^E_j$. (c) Equilibrium prices $p^E_i$ and $p^E_j$. }
\label{equilibrium_figure}
\end{figure*}

\vskip 10pt \noindent {\bf Proposition 3}: {\it Under the regulation
that limits the price level changes by MNOs, if the unit cost per
capacity $\gamma$ satisfies the following inequality,
\setlength\arraycolsep{1pt} \begin{eqnarray}  \label{eq:gamma} F
\left( \gamma \right )=\sqrt[3]{{\frac{2}{\gamma } + \sqrt
{\frac{4}{{\gamma ^2 }} + \frac{1}{{27\gamma ^3 }}} }} +
\sqrt[3]{{\frac{2}{\gamma } - \sqrt {\frac{4}{{\gamma ^2 }} +
\frac{1}{{27\gamma ^3 }}} }} > 2, \nonumber \\
\end{eqnarray}
then there is an equilibrium point $(k^E_i,k^E_j,p^E_i,p^E_j)$:
\setlength\arraycolsep{0pt} \begin{eqnarray} &&\left(
k^E_i,k^E_j,p^E_i,p^E_j \right) \nonumber \\ &=&\left( {k_i^E ,\sqrt
{\frac{{k_i^E + 1}}{{\left( {k_i^E  + 2} \right)^2 \gamma }}}  -
1,\frac{1}{{k_i^E + 2}},\frac{{k_i^E  + 1}}{{k_i^E  + 2}}} \right)
\label{eq:market_power} \nonumber \\ &=&\left ( F \left( \gamma
\right )-2, \sqrt {\frac{{F \left( \gamma \right )-1}}{{F \left(
\gamma \right )^2 \gamma }}}  - 1 ,{\frac{1}{{F \left( \gamma \right
)}},1-\frac{1}{{F \left( \gamma \right )}}} \right ), \nonumber \\
\end{eqnarray}
which is Pareto-optimal.}

\vskip 10pt \noindent {\bf {Proof}}: In Case 2 $(k_i^* \ge 2k_j^*)$,
to be an equilibrium point, the optimal solution should satisfy the
following equation:
\begin{eqnarray}
k_i^*  &=& \max \left\{ {0,\sqrt {\frac{{2k_j^*  + 1}}{{4\left(
{k_j^* + 1} \right)^2 \gamma }}}  - 1} \right\} \nonumber \\ &\le&
\max \left\{ {0,\sqrt {\frac{{k_j^{*2}  + 2k_j^*  + 1}}{{4\left(
{k_j^* + 1} \right)^2 \gamma }}}  - 1} \right\} \nonumber \\ &=&
\max \left\{ {0,\sqrt {\frac{1}{{4\gamma }}}  - 1} \right\} = k_j^*,
\end{eqnarray}
which contradicts the assumption $k_i^* \ge 2k_j^*$ because $k_i^*$
and $k_j^*$ are non-zero. Therefore, there is no equilibrium point
in Case 2.

In Case 1 $(k_i^* \le 2k_j^*)$, to be an equilibrium point, the
optimal solution should satisfy the following equation:
\begin{eqnarray}
&& 2k_j^*  - k_i^* = 2\sqrt {\frac{{k_i^*  + 1}}{{\left( {k_i^*  +
2} \right)^2 \gamma }}}  - \left( {k_i^*  + 2} \right) \ge 0
\nonumber \\ &&\Rightarrow  2\sqrt {\frac{{k_i^* + 1}}{{\left(
{k_i^* + 2} \right)^2 \gamma }}} \ge  {k_i^* + 2} .
\end{eqnarray}
The left and right hand side equations of the last inequality are
positive. Thus, we compare the squares of them as follows:
\setlength\arraycolsep{1pt}
\begin{eqnarray}
&& \left( 2\sqrt {\frac{{k_i^*  + 1}}{{\left( {k_i^*  + 2} \right)^2
\gamma }}}  \right)^2  - \left( {k_i^*  + 2} \right)^2 \nonumber
\\ &=& \frac{{4k_i^*  + 4 - \left( {k_i^*  + 2} \right)^4 \gamma
}}{{\left( {k_i^*  + 2} \right)^2 \gamma }}  \nonumber \\ &=&
\frac{{4k_i^*  + 4 - \left( {k_i^*  + 2} \right)\left( { - k_i^*  +
2} \right)}}{{\left( {k_i^* + 2} \right)^2 \gamma }}  \nonumber \\
&=& \frac{{k_i^* \left( {k_i^* + 4} \right)}}{{\left( {k_i^*  + 2}
\right)^2 \gamma }} \ge 0. \nonumber
\end{eqnarray}
Note that the second equality holds because we use an equation
$(k_i^*+2)^3\gamma=-k_i^*+2$ from the first order condition of the
optimization problem. From the above calculations, we conclude that
if $k_i^*>0$ (i.e., $F(\gamma)>2$), then there would be an
equilibrium. Using these results and Lemma 4, we can calculate the
equilibrium as in Proposition 3. \hfill $\blacksquare$

\vskip 10pt We call $F(\gamma)$ of (\ref{eq:gamma}) the {\it
feasibility function} because we can discriminate between feasible
and infeasible markets using this function.\footnote{By a feasible
market, we mean there is at least one operator wishing to exist for
the market.} If the unit cost $\gamma$ is expensive and does not
satisfy Equation (\ref{eq:gamma}), then the market will be
infeasible (i.e., market failure) and, $k_i^E$ and $k_j^E$ are
negative values. Note the equilibrium point is a function of $k^E_i$
in Equation (\ref{eq:market_power}). This means MNO $i$ has {\it
market power} even though the regulator gives the last opportunity
for changing the price level to MNO $j$.

Figure \ref{equilibrium_figure}-(a) shows the feasibility function,
where the unit cost $\gamma$ should be less than $0.25$ to avoid
market failure. We plot the equilibrium point in Proposition 3
varying $\gamma \in [0,0.25]$. Figures \ref{equilibrium_figure}-(b)
and \ref{equilibrium_figure}-(c) show the result. We observe that
$p^E_j$ is always higher than $p^E_i$, and MNO $j$ always invests
more than MNO $i$. This means the users with high user type (i.e.,
high QoS requirement and high willingness-to-pay) are targeted by
MNO $j$. On the other hand, MNO $i$ can make a profit by making a
relatively small investment because its target users have low user
types. An interesting observation is that the price gap between both
MNOs decreases as the unit cost per capacity increases in Figure
\ref{equilibrium_figure}-(c). This is because the high cost makes
both MNOs reduce their investment levels and concentrate on
lucrative targets (i.e., the users whose user types are near $0.5$).

%\begin{figure}[t]
%\centerline{\epsfig{figure=Fig7.eps,height=2.8in,clip=;}}
%\caption{Feasibility function $F(\gamma)$.}
%\label{feasibility_function}
%\end{figure}
%
%\begin{figure}[t]
%\centerline{\epsfig{figure=Fig8.eps,height=2.8in,clip=;}}
%\caption{Equilibrium capacities $k^E_i$ and $k^E_j$.}
%\label{equilibrium_price_figure}
%\end{figure}
%
%\begin{figure}[t]
%\centerline{\epsfig{figure=Fig9.eps,height=2.8in,clip=;}}
%\caption{Equilibrium prices $p^E_i$ and $p^E_j$.}
%\label{equilibrium_capacity_figure}
%\end{figure}

\subsection{Role of the Regulator}
\begin{figure}[t]
\centerline{\epsfig{figure=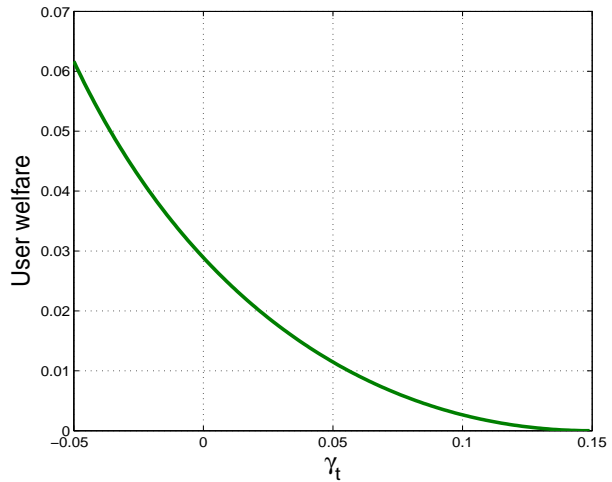,width=3.5in,
height=2.7in,clip=;}} \caption{User welfare as a function of
$\gamma_t$ ($\gamma_c=0.1$). The value is divided by $M$.}
\label{user_number}
\end{figure}

The regulator's key concern is to improve user welfare \cite{Wpin},
\cite{KimYu}. User welfare means the sum of all users' utilities. If
a user with user type $\theta$ purchases MNO $i$'s network service,
its net utility will be $\theta - p_i$. On the other hand, if the
user consumes neither of MNOs' network services, then its utility
will be zero. The regulator can achieve its purpose by exacting {\it
taxes} from the MNOs or giving {\it subsidies} to them. So far, we
assume that the unit cost per capacity $\gamma$ is a given
parameter. However, we can divide $\gamma$ into $\gamma_c$ and
$\gamma_t$ (i.e., $\gamma=\gamma_c+\gamma_t$). The value $\gamma_c$
denotes the fixed cost, and $\gamma_t$ denotes the tax ($\gamma_t
> 0$) or subsidy ($\gamma_t < 0$). We plot user welfare as a function of $\gamma_t$ in Figure \ref{user_number}. In the figure, we
set $\gamma_c=0.1$. The figure shows that user welfare decreases as
$\gamma_t$ increases. Even though this result is predictable, the
regulator can use it to forecast results of exacting taxes or giving
subsidies.

\begin{figure}[t]
\centerline{\epsfig{figure=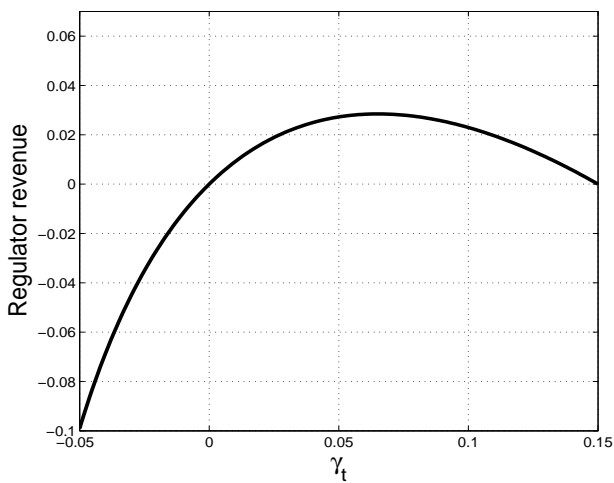,width=3.5in,
height=2.7in,clip=;}} \caption{The regulator's revenue as a function
of $\gamma_t$ ($\gamma_c=0.1$). The value is divided by $M$.}
\label{regulator_revenue}
\end{figure}

From the regulatory perspective, another important thing is to
secure finances. We plot the regulator's revenue as a function of
$\gamma_t$ in Figure \ref{regulator_revenue}. Intuitively, the
regulator runs a deficit when it gives subsidies (i.e.,
$\gamma_t<0$) to improve user welfare. Therefore, the regulator
should strike a balance between securing finances and improving user
welfare. Figure \ref{regulator_revenue} also shows that very high
taxes lead to the revenue loss. This is because the burden of high
taxes makes MNOs cut their investments. If the regulator's goal is
to maximize its revenue, then $\gamma_t$ should be set to $0.065$.

%\begin{itemize}
%\item {\it Fairness perspective}:  We plot the profits of both MNOs in Figure \ref {revenue}. The figure shows that the profit of
%MNO $i$ decreases as $\gamma$ increases. Interestingly, the profit
%of MNO $j$ increases as $\gamma$ increases. This is because MNO $i$
%should reduce its investment amount when $\gamma$ is high, which
%gives MNO $j$ an opportunity to expand its service to the users of
%low user type who could not access MNO $j$'s network before. The
%regulator can adjust the fairness between the MNOs by exacting taxes
%or giving subsidies. In the figure, we observe that $\gamma$ should
%be about $0.13$ to ensure the fairness.
%\end{itemize}
%
%\begin{figure}[t]
%\centerline{\epsfig{figure=Fig11.eps,height=2.8in,clip=;}}
%\caption{Profits of both MNOs. The values are divided by $M$.}
%\label{revenue}
%\end{figure}

\section{Conclusions}
MNOs tend to compete with each other changing their service prices
by subsidization in the real world (see Figure \ref{subsidy}). In
this paper, to theoretically explain the price dynamics in the
mobile communication service, we used a two-stage Cournot and
Bertrand competition model that is well understood in
microeconomics. The Cournot and Bertrand models are interlinked and
we perform a joint optimization of network capacity and service
price. Based on a game-theoretic approach, we show that there is a
price war with long jumps. This price dynamics explains the subsidy
dynamics in the real world. To avoid the instability and
inefficiency, we propose a regulation that ensures an equilibrium
point of price levels, which is Pareto-optimal. Based on our results
in the Cournot stage, we describe characteristics of the duopoly
market and suggest the regulator's optimal actions (exacting taxes)
corresponding to user welfare and the regulator's revenue. Although
our analytic results are derived under some assumptions for
mathematical tractability, it will provide good intuition for
understanding the price dynamics and imposing regulations in the
mobile communication service.

\end{document}